\documentclass[11pt,aps,showpacs,nofootinbib,longbibliography]{revtex4-1}

\pdfoutput=1

\usepackage{ulem}
\usepackage{siunitx}
\usepackage{comment}
\usepackage[dvipsnames]{xcolor}
\usepackage{amssymb,amsmath,latexsym,mathrsfs}
\usepackage[USenglish,american]{babel}
\usepackage{bm}
\usepackage{amsfonts}
\usepackage{graphicx}
\usepackage{epstopdf}
\usepackage{hyperref}
\usepackage{array}
\usepackage[utf8]{inputenc}
\usepackage{soul}
\usepackage{color}
\usepackage{slashed}
\usepackage{csquotes}
\usepackage[T1]{fontenc}

\usepackage{footnote}

\everymath{\displaystyle} 

\usepackage{soul}

\newcommand{\beq}{\begin{equation}}
\newcommand{\eeq}{\end{equation}}
\newcommand{\beqa}{\begin{eqnarray}}
\newcommand{\eeqa}{\end{eqnarray}}
\newcommand{\bea}{\begin{eqnarray}}
\newcommand{\eea}{\end{eqnarray}}

\newcommand{\rr}{f}

\begin{document}

\title{Core and Halo Properties in
Multi-Field Wave Dark Matter}

\author{Fabio van Dissel$^{1}$, Mark P. Hertzberg$^{2,3,4}$, Jared Shapiro$^{2}$} 
\email{\baselineskip 11pt Email addresses: fvdissel@ifae.es; mark.hertzberg@tufts.edu; jared.shapiro@tufts.edu}

\affiliation{ $^1$Institut de F\'{\i}sica d'Altes Energies (IFAE)\\ 
 Campus UAB, 08193 Bellaterra (Barcelona) Spain\\
$^2$Institute of Cosmology, Department of Physics and Astronomy\\
Tufts University, Medford, MA 02155, USA\\
$^3$Institute of Theory and Computation, Center for Astrophysics \\
Harvard University, Cambridge, MA 02138, USA\\
$^4$Center for Theoretical Physics, Department of Physics\\
Massachusetts Institute of Technology, Cambridge, MA 02139, USA
 }

\begin{abstract}
In this work, we compute multi-field core and halo properties in wave Dark Matter models. We focus on the case where Dark Matter consists of two light (real) scalars, interacting gravitationally. As in the single-field Ultra Light Dark Matter (ULDM) case, the scalar field behaves as a coherent BEC with a definite ground state (at fixed total mass), often referred to in the literature as a gravitational soliton. We establish an efficient algorithm to find the ground and excited states of such two-field systems. We then use simulations to investigate the gravitational collapse and virialization, starting from different initial conditions, into solitons and surrounding halo. As in the single-field case, a virialized halo forms with a gravitational soliton (ground state) at the center. We find some evidence for an empirical relation between the soliton mass and energy and those of the host halo. We use this to then find a numerical relation between the properties of the two. 
Finally, we use this to address the issue of alleviating some of the tensions that single-field ULDM has with observational data, in particular, the issue of how a galaxy's core and radius are related.
We find that if galaxies of different masses have similar percentages of the two species, then the core-radius scaling tension is not addressed. However, more general possibilities occur if the relative abundance of species in each halo correlates with the total mass of the galaxy. If this is the case, the model predicts several other phenomenological signatures. 

\end{abstract}
\maketitle

\newpage 

\tableofcontents

\section{Introduction}
\label{sec:intro}

The nature of dark matter remains an important open question. One of the leading candidates for dark matter is a very light boson. For this candidate to make up all the dark matter in the universe, they must have a large number density in the galactic halos. In this case, the particle's occupancy number can be very large, meaning that the theory can be approximated by classical field theory. These classical fields enjoy wave-like properties, such as interference, etc. 

One motivation comes from the idea of ultra-light axions, whose presence can be accommodated in fundamental physics \cite{Arvanitaki_2010}. In this case, it is plausible that the particles have a mass on the order of $m\sim 10^{-21}\,$eV, or so. Then, the particle's de Broglie wavelength in a galaxy can be huge. Since typical virial velocities in galaxies are $v\sim 10^{-5}-10^{-3}$\,c, the corresponding de Broglie wavelength $\lambda_{dB}\sim h/(mv)$ can be on the order of kpc, or so. Such large de Broglie wavelengths smooth out the centers of galaxies, producing a core rather than a cusp, and can reduce small scale structure generally \cite{Hu:2000ke}. For many years this has been suggested as a feature, as some of the simplest simulations of dark matter show a moderately spiky feature near the center, rather than that seen in observations. Whether this so-called ``core-cusp problem" is real or not remains a matter of debate. Nevertheless it has historically provided one motivation to consider the phenomenological consequences of ultra-light bosonic dark matter. Furthermore, the idea of ultra-light bosons as dark matter is interesting in its own right as it provides new wave-phenomenology, such as interference patterns, not present in standard (heavy) dark matter models.

Despite these interesting motivations, in the last few years, observational constraints on single component Ultra Light Dark Matter (ULDM) have become more and more severe \cite{dalal2022fuzzy, Ir_i__2017, Armengaud_2017, Chan_2022, Powell_2023,Hertzberg:2022vhk}. There is therefore increasing interest in opening up more parameter space in the ULDM picture by adding more light scalars to the model \cite{Huang_2023, luu2023nested, Gosenca_2023, Glennon_2023, jain2023kinetic, Eby_2020, Amin_2022}. 
For example, this can suppress the heating of stars near cores, suppressing the effect mentioned in Ref.~\cite{dalal2022fuzzy}.
Moreover, multiple light scalars is often argued as more natural from the point of view of fundamental physics (e.g., \cite{Arvanitaki_2010}). 

In this work, we will pay particular attention to another apparent tension that exists between single component ULDM and observational data; 
there are hints from data on galaxies (e.g., see Ref.~\cite{Rodrigues_2017}) that the size of the galactic core $R_c$ and the corresponding density $\rho_c$ obey an approximate scaling relation $\rho_c\propto 1/R_c^\beta$, with $\beta\sim 1$ (in fact the value $\beta\approx1.3$ is a best fit value to a set of galaxies. However, in Ref.~\cite{Deng_2018} it was shown that this scaling relation is not compatible with any single-field bosonic model. In the simple Newtonian gravity dominated regime, the single-field bosonic model predicts a relation $\rho\propto 1/R_c^\beta$ with $\beta=4$. Furthermore, Ref.~\cite{Deng_2018} showed that if a self-interacting potential $V$ is included, there is no choice that leads to the observed scaling with stable solutions.

In Ref.~\cite{Guo_2021}, it was suggested that a two-field bosonic model may help the situation. Here it is explained that the space of solutions is increased, leading to a larger array of possibilities than the precise $\rho_c\propto 1/R_c^4$ prediction of the single field model in the Newtonian gravity dominated regime.

In this work, we wish to improve upon this work on multi-field models in several ways: (i) we will calculate the space of solitonic solutions more carefully, (ii) we introduce a prescription that allows one to automate the solitonic solution for multi-field models, (iii) we run simulations (albeit within a spherically symmetric restriction) to obtain scaling relations, (iv) we learn the trends of the space of solitons that naturally arise from different kinds of initial conditions, (v) we lay out the requirements in how the relative fraction of fields must occupy different galaxies in order to better explain the data, or else, to falsify the proposal.

Our paper is organized as follows: 
In Section \ref{sec:SPs} we start with the relativistic theory of two-scalar fields, and take the nonrelativistic limit, and describe the basic solitonic (static) solutions.
In Section \ref{sec:Dynamical} we formulate a numerical dynamical treatment, from different choices of initial conditions, to determine which types of solitons emerge. We obtain some empirical scaling relations from these results, albeit the validity of this scaling needs to be subjected to larger scrutiny in future work. 
In Section \ref{sec:cosm} we present possible cosmological implications of these results, through establishing a multi-field relation between the soliton's properties and the halo mass.
In Section \ref{sec:Conclusion} we conclude and discuss our findings.
Finally, in the Appendices, we provide more details on the phenomenology of the model.
%non-relativistic limit and the 

\section{The Schr\"odinger-Poisson system}
\label{sec:SPs}
We consider two scalar fields minimally coupled to gravity with action (signature -+++, units $\hbar=c=1$):
\beq
S = \int d^4x \sqrt{-g} \left[\frac{\mathcal{R}}{ 16 \pi G} + \mathcal{L_M}\right]
\label{eq:action}
\eeq
Where the matter contribution to the action is provided by two massive (real) scalars
\beq
\mathcal{L_M} = -\frac{1}{2}\partial_\mu \phi_1 \partial^\mu\phi_1 - \frac{1}{2}\partial_\mu \phi_2 \partial^\mu\phi_2 - \frac{1}{2} m_1^2 \phi_1^2 - \frac{1}{2} m_2^2 \phi_2^2
\label{eq:langmatter}
\eeq
Where $m_1$ and $m_2$ are the masses of $\phi_1$ and $\phi_2$ respectively. We can also add higher order nonlinear terms in a potential, but this is the leading-order terms for any two-scalar system. Terms proportional to higher powers of $\phi$ are in general also present, in particular if the scalar under consideration is an axion-like particle, self-interactions should be present from a cosine-like potential. However, in the current work we will assume to be in a regime where we can safely neglect those terms and focus on the dynamics of the system given in Eqs.~\eqref{eq:action} and \eqref{eq:langmatter}. Since we are concerned with Dark Matter, we can assume the typical velocities of the scalar particles to be of order $v \sim 10^{-5} - 10^{-3} c$, which is the typical velocities of particles in a galactic halo (from dwarfs to large galaxies). Therefore, we can safely move to the non-relativistic regime, in which we treat gravity in the Newtonian limit.   

\subsection{The Non-Relativistic Limit of Scalar Fields}
The non-relativistic limit of the action in Eq.~\eqref{eq:action} has been extensively discussed in the literature. We present a brief derivation here. %, as more extensive analyses already exist in the literature. 
%a more extensive review for Appendix~\ref{sec:appendixA}. 
The basic procedure is to decompose each real scalar in a sum of complex degrees of freedom, after factorizing out its primary oscillations in an $e^{-imt}$ factor, as follows
\bea
\phi_1 = \frac{1}{\sqrt{2 m_1}}\left(\psi(\vec{x}, t) e^{-i m_1 t} + c.c.\right)\\
\phi_2 = \frac{1}{\sqrt{2 m_2}}\left(\chi(\vec{x}, t) e^{-i m_2 t} + c.c.\right)
\eea
The non-relativistic limit then corresponds to assuming that the complex degrees of freedom $\psi$ and $\chi$ are slowly varying in space and time. Relatedly, we preserve the degrees of freedom by building an action that involves only {\em one} derivative acting on $\psi,\,\chi$, rather than the two derivatives acting on $\phi_1,\,\phi_2$. We are in the limit where $|\nabla \psi| \ll m_1 |\psi|$ and $|\dot{\psi}| \ll m_1 |\psi|$.  Similarly $|\nabla \chi| \ll m_2 |\chi|$ and $|\dot{\chi}| \ll m_2 |\chi|$. Lastly, we assume to be in a weak field limit where $\frac{|\psi|}{\sqrt{m_1} M_{Pl}} \sim\frac{|\chi|}{\sqrt{m_1} M_{Pl}}  \ll 1$ ($M_{Pl}=1/\sqrt{G}$). In this limit, gravity becomes Newtonian, with the only relevant term in the metric being $g_{tt} = -1 - 2 U$, where $U$ is the gravitational potential. The Lagrangian density then becomes (to lowest dynamical order)
\bea
\mathcal{L}_{nr} &=& -\frac{\nabla U \nabla U}{8 \pi G} - \frac{1}{2 m_1} \nabla \psi \nabla \psi^* - \frac{1}{2 m_2} \nabla \chi \nabla \chi^* - U \left(m_1 |\psi|^2 + m_2 |\chi|^2\right) \nonumber\\
&&+ \frac{i}{2}\left(\dot{\psi} \psi^* - \dot{\psi^*} \psi \right) + \frac{i}{2}\left(\dot{\chi} \chi^* - \dot{\chi^*} \chi \right) 
\label{eq:nrlagrangian}
\eea

Interestingly, in this non-relativistic limit, there is an accidental pair of $U(1)$ global symmetries in Eq.~\eqref{eq:nrlagrangian} (the action is invariant under a change of the field's phase). So the system contains two conserved quantities $N_1=\int d^3x\, n_1(\vec{x},t)$ and $N_2=\int d^3x\, n_2(\vec{x},t)$, with corresponding number densities $n_1(\vec{x},t) = |\psi|^2$ and $n_2(\vec{x},t) = |\chi|^2$. These are just the number densities of the particles of the two respective species (the mass density is then given by $\rho_i(\vec{x}) = m_i n_i(\vec{x})$), which is conserved in the non-relativistic regime as particle number changing processes are relativistic and suppressed. 

The expressions for the masses and energies in each field are
\beq
M_\psi = \int d^3x \,m_1 |\psi|^2 \quad \textrm{and} \quad M_\chi = \int d^3x\,  m_2 |\chi|^2
\eeq
\beq
K_\psi = \int d^3x\,  \frac{1}{2 m_1} \nabla \psi \nabla \psi^* \quad \textrm{and} \quad K_\chi = \int d^3x\,  \frac{1}{2 m_2} \nabla \chi \nabla \chi^*
\eeq
\beq
W_\psi = \int d^3x\,  \frac{m_1 U}{2} |\psi|^2\quad \textrm{and} \quad W_\chi = \int d^3x\,  \frac{m_2 U}{2} |\chi|^2
\eeq
With $M_i$, $K_i$ and $W_i$ the mass, kinetic and gravitational energy of each field respectively. The total energy of the system is given by (neglecting the rest mass energy which always dominates in the NR-regime) 
\beq
E_{tot}=K_\psi+K_\chi+W_\psi+W_\chi
\eeq

By varying the above action, the dynamics are described by the well-known Schrödinger-Poisson system of equations
\bea
i \dot{\psi} &=& - \frac{\nabla^2}{2 m_1} \psi + m_1 U \psi
\label{eq:SP1}\\
i \dot{\chi} &=& - \frac{\nabla^2}{2 m_2} \chi + m_2 U \chi
\label{eq:SP2}
\eea
where the gravitational potential $U$ is the solution to the Poisson equation
\beq
\nabla^2 U = 4 \pi G\left(m_1 |\psi|^2 + m_2 |\chi|^2\right)
\label{eq:SP3}
\eeq
This set of equations will form the basis of our study, which to first order, should describe the dynamics of scalar dark matter particles on galactic scales. Since we will exploit it later, we note that there is a scale transformation that leaves this set of equations invariant. In particular,
\beq
\psi(\vec{x}, t) \rightarrow \lambda^2\, \psi(\lambda \vec{x}, \lambda^2 t)
\label{eq:psitrans}
\eeq
\beq
\chi(\vec{x}, t) \rightarrow \lambda^2\, \chi(\lambda \vec{x}, \lambda^2 t)
\label{eq:phitrans}
\eeq
\beq
U(\vec{x}) \rightarrow \lambda^2\, U(\lambda \vec{x})
\label{eq:newttrans}
\eeq
Where in Eq.~\eqref{eq:newttrans} we suppress the dependence of $U$ on $t$ as the gravitational potential is itself nondynamical, and purely sourced by the presence of the scalar fields. 

Eqs.~\eqref{eq:SP1} \eqref{eq:SP2} and \eqref{eq:SP3}  describe the behavior of a coherent Bose-Einstein condensate, with conserved particle numbers $N_1$ and $N_2$. To understand the dynamics of the system, it is therefore quintessential to find its static solutions at fixed particle number, akin to the eigenstates of the Hamiltonian in quantum mechanics. 

\subsection{Static Solutions}
\label{sec:staticsols}
The static solutions of the Schrödinger-Poisson system are those solutions whose gravitational potential remains constant over time. Therefore the time-dependence of the scalar has to be a pure space-independent phase. In particular, one can look for solutions of the form
\beq
\psi(\vec{x}, t) = \left(\frac{\sqrt{m_1} M_{Pl}}{\sqrt{4 \pi}}\right) \Psi(r) e^{-i \gamma m_1 t}
\label{eq:dimensionlessphi1}
\eeq
\beq
\chi(\vec{x}, t) = \left(\frac{\sqrt{m_1} M_{Pl}}{\sqrt{4 \pi}}\right) \Phi(r) e^{-i \kappa m_2 t}
\label{eq:dimensionlessphi2}
\eeq
Where $\gamma$ and $\kappa$ are chemical potentials for each of the species. We limit our search to spherical solutions, as those are the ones that generally will minimize the energy of the configuration. We can take the radial profiles, $\Psi$ and $\Phi$, to be real functions without loss of generality.

The problem is thus reduced to solving the eigenvalue problem characterized by the following set of equations, where we introduced dimensionless variables through $\tilde{x}_\mu = m_1 x_\mu$ 
\bea
\nabla^2 \Psi &=& 2 \left(U - \gamma\right) \Psi
\label{eq:SP1static}\\
\nabla^2 \Phi &=& 2 \left(\frac{m_2}{m_1}\right)^2 \left(U - \kappa\right) \Phi
\label{eq:SP2static}
\eea
with the gravitational Poisson equation
\beq
\nabla^2 U = \left(\Psi^2 + \frac{m_2}{m_1} \Phi^2\right)
\label{eq:SP3static}
\eeq
where in spherical symmetry the Laplacian is 
\beq\nabla^2 = \partial^2_r + \frac{2}{r}\partial_r
\eeq

We are looking for localized solutions of the above set of equations. We first fix the central amplitude of the fields to $\Psi(0) = \Psi_0$ and $\Phi(0) = \Phi_0$. Next, imposing the regularity conditions $\partial_r\Psi(0) = \partial_r\Phi(0) = \partial_rU(0) = 0$, the system can effectively be solved through a two-parameter shooting method. The ground state solutions are those in which the fields have no nodes, and monotonically approach zero at large $r$.

\subsection{Iterative Procedure}
In the single field case, there is only one chemical potential and the shooting method is rather straightforward (work on the single field case includes Refs.~\cite{Chavanis:2011zi,Chavanis:2011zm,Schiappacasse:2017ham,Visinelli:2017ooc}). However,
we found that it's difficult to find the localized solution, while shooting the two parameters $\gamma$ and $\kappa$ simultaneously. We thus employ an iterative method which we found to converge efficiently. First, we split the gravitational potential into two parts, $U(r) = U_\Psi(r) + U_\Phi(r)$, where
\bea
&&\nabla^2 U_\Psi = \Psi^2
\label{eq:Upsi1}\\
&&\nabla^2 U_\Phi = \frac{m_2}{m_1} \Phi^2
\label{eq:Uphi1}
\eea
The iterative method can then be summarized as
\begin{enumerate}
    \item Initially, set $U_\Phi = 0$ and find the localized solution of the system
        \beq
        \nabla^2 \Psi = 2 \left(U_\Psi + U_\Phi - \gamma\right) \Psi
        \label{eq:SP1iter1}
        \eeq
        \beq
        \nabla^2 U_\Psi = \Psi^2
        \label{eq:SP2iter1}
        \eeq
        (even though in this very first step $U_\Phi=0$, we formally include it in the first equation here, as this will be important when we repeat the procedure, which will involve a nonzero $U_\Phi$).
    This system can be solved straightforwardly using shooting methods familiar from the single field case as it involves only one parameter $\gamma$.
    \item Updating the solution of $U_\Psi$ obtained from step 1, find the localized solution of the system
        \beq
        \nabla^2 \Phi = 2 \left(\frac{m_2}{m_1}\right)^2  \left(U_\Psi + U_\Phi - \kappa\right) \Phi
        \label{eq:SP1iter2}
        \eeq
        \beq
        \nabla^2 U_\Phi = \frac{m_2}{m_1} \Phi^2
        \label{eq:SP2iter2}
        \eeq
    Again, there is no difficulty in finding the localized solutions of this system as it involves only one parameter $\kappa$.
    \item Iterate through steps 1 and 2, making sure to update the gravitational potentials at each step until convergence of the solutions is reached. We define convergence by the two eigenvalues $\gamma$ and $\kappa$. In practice, we see that these two parameters don't change after a certain number of iterations after which we stop the procedure.
\end{enumerate}
In practice, we find that this method converges after $\mathcal{O}(10)$ iterations. Note that this method can straightforwardly be extended to include non-gravitational interactions between the scalars, such as quartic interactions. However, for realistic models, such interactions are normally negligible, and in any case, is beyond the scope of the current work. 

At fixed central density $\Psi_0$ and $\Phi_0$, there are an enumerable infinite amount of solutions labeled by $n_1$ and $n_2$, indicating the number of nodes in $\Psi$ and $\Phi$ respectively. To find any particular solution we have to solve for $\Psi$ and $\Phi$ with the appropriate number of nodes through each iteration. In the remainder of this work, we will focus on the ground state (no nodes) configurations with $n_1 = n_2 = 0$, as it is expected that those are the gravitational solitons that form at the center of galaxies. 
We are thus able to find the static ground state solutions of the Schrödinger-Poisson system of equations at fixed central densities $\Psi_0$, $\Phi_0$. 

The scaling relations of Eqs.~\eqref{eq:psitrans}, \eqref{eq:phitrans} and \eqref{eq:newttrans} then allow us to find any solution that has the same ratio of central densities $\rr = \frac{\Phi_0}{\Psi_0}$, as this ratio is conserved through the transformation. Therefore, it is only necessary to find the ground state soliton once for each ratio of central densities, noting also that if the ratio is extreme, the ground state effectively becomes the single-field soliton. The properties of the different solutions are related to each other via
\bea
&&M_{\Psi,\rr,\lambda} = \lambda M_{\Psi,\rr,0}\quad \textrm{and} \quad M_{\Phi,\rr,\lambda} = \lambda M_{\Phi,\rr,0}
\label{eq:masstrans}\\
&&K_{\Psi,\rr,\lambda} = \lambda^3 K_{\Psi,\rr,0}\quad \textrm{and} \quad K_{\Phi,\rr,\lambda} = \lambda^3 K_{\Phi,\rr,0}
\label{eq:kintrans}\\
&&W_{\Psi,\rr,\lambda} = \lambda^3 W_{\Psi,\rr,0} \quad \textrm{and} \quad W_{\Phi,\rr,\lambda} = \lambda^3 W_{\Phi,\rr,0}
\label{eq:pottrans}
\eea
Where the subscript $\rr,0$ corresponds to a reference value at each ratio $\rr$. In what follows we will take the reference solutions to be the solutions where $\Psi_0 = 1$ and $\Phi_0 = \rr$. 

\begin{figure}[t]
    \centering
    \includegraphics[width=\textwidth]{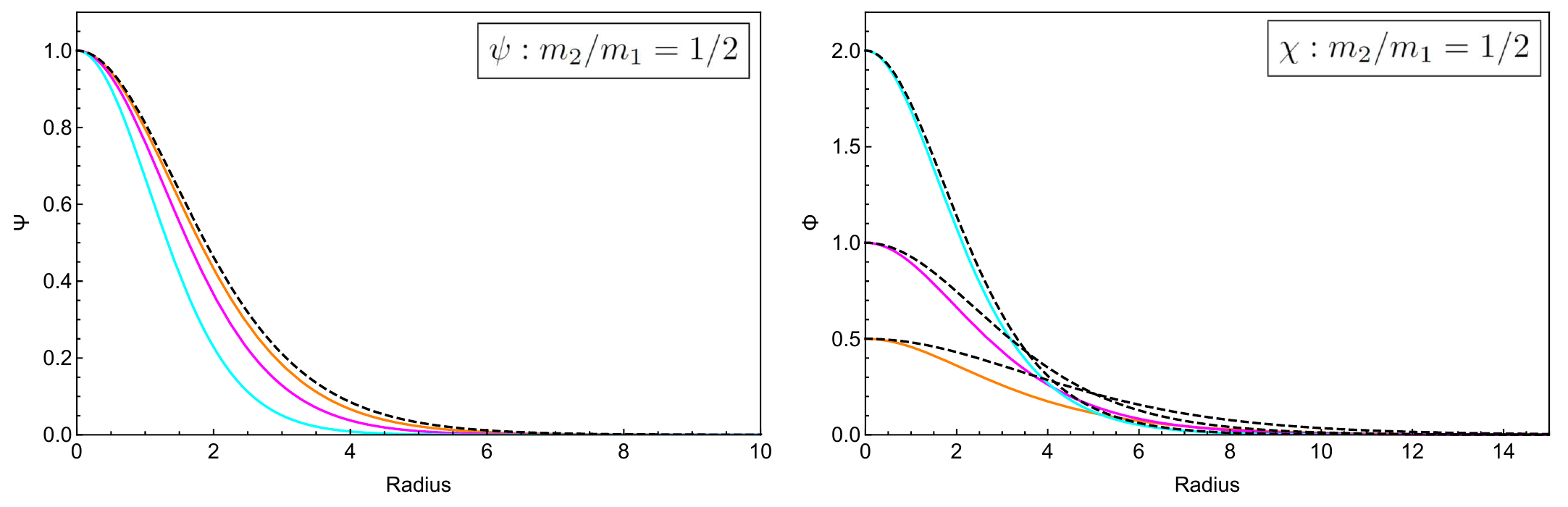}
     \includegraphics[width=\textwidth]{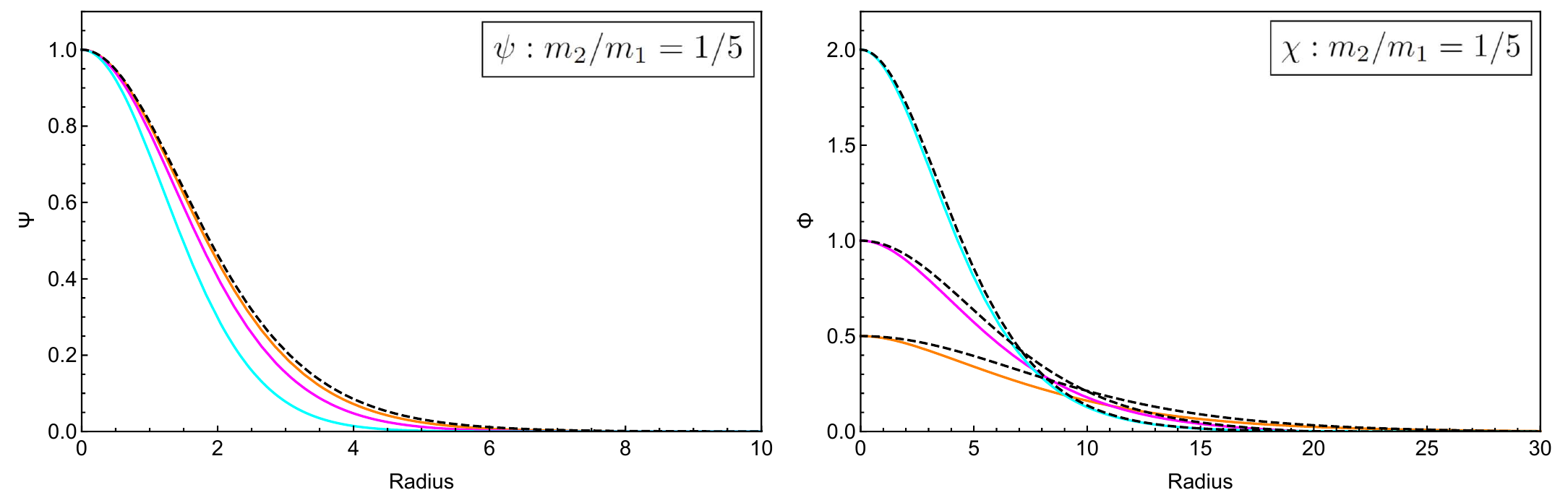}
    \caption{{\it Top panel:} Some ground state soliton solutions in the case where $m_2/m_1 = 1/2$. 
    {\it Bottom panel:} Some ground state soliton solutions in the case where $m_2/m_1 = 1/5$.
    Here we plot three cases: one where the heavier field dominates the central density contribution (orange), an intermediate case (magenta), and a case where the light field dominates (cyan). The black dashed line corresponds to the single-field solution which matches best in the case where one of the two fields dominates the total mass of the solution. {\it Left:} The field profile of the heavy field $\Psi(r)$. {\it Right:} The field profile of the light field $\Phi(r)$}
    \label{fig:groundstatesolutionsm20d5}
\end{figure}

\begin{figure}[h]
    \centering
    \includegraphics[width=\textwidth]{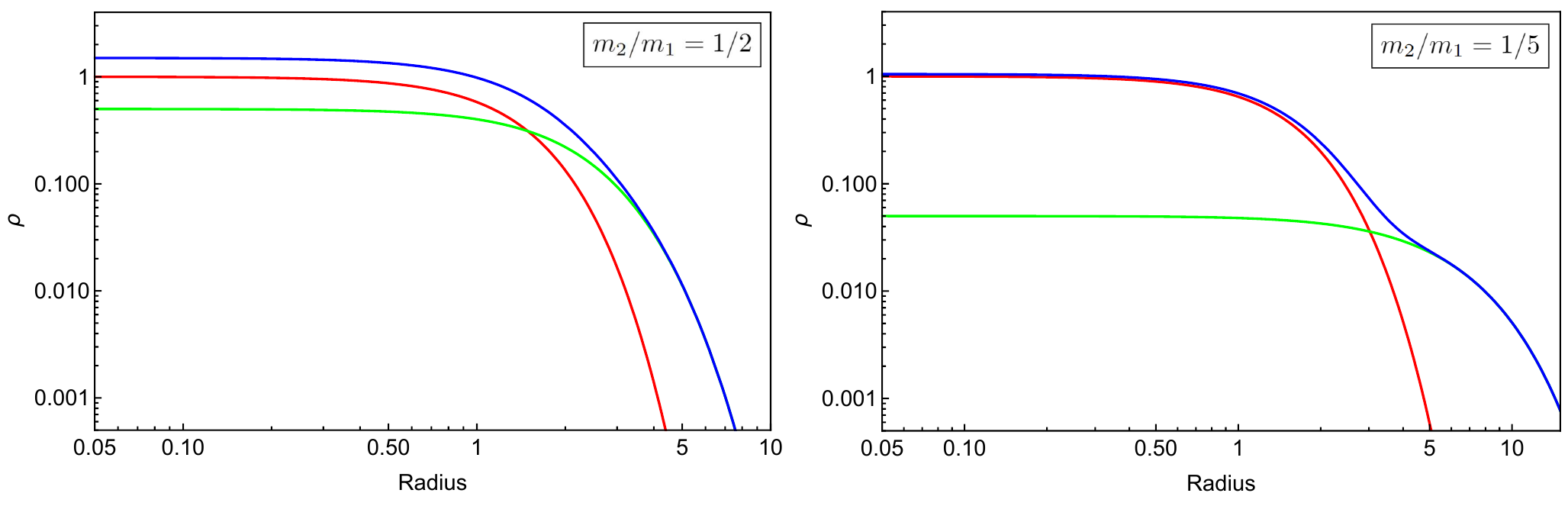}
    \caption{The mass density profile for a typical ground state soliton solution. We plot the mass density $\rho$ for the heavy field (red), light field (green), and the total sum (blue). The nested nature of the soliton is clearly visible, where initially the heavy field dominates the total density near the center. However, due to the larger De Broglie wavelength of the lighter field, there exists a "turnover point" where the light field starts to dominate. {\it Left:} $m_2/m_1 = 1/2$ corresponding to the magenta profiles of Fig.~\ref{fig:massdensityprofs}. {\it Right:} $m_2/m_1 = 1/5$ corresponding to the orange profiles of Fig.~\ref{fig:massdensityprofs}.}
    \label{fig:massdensityprofs}
\end{figure}

\subsection{Sample Solutions}
\begin{figure}[h]
    \centering
    \includegraphics[width=0.6\textwidth]{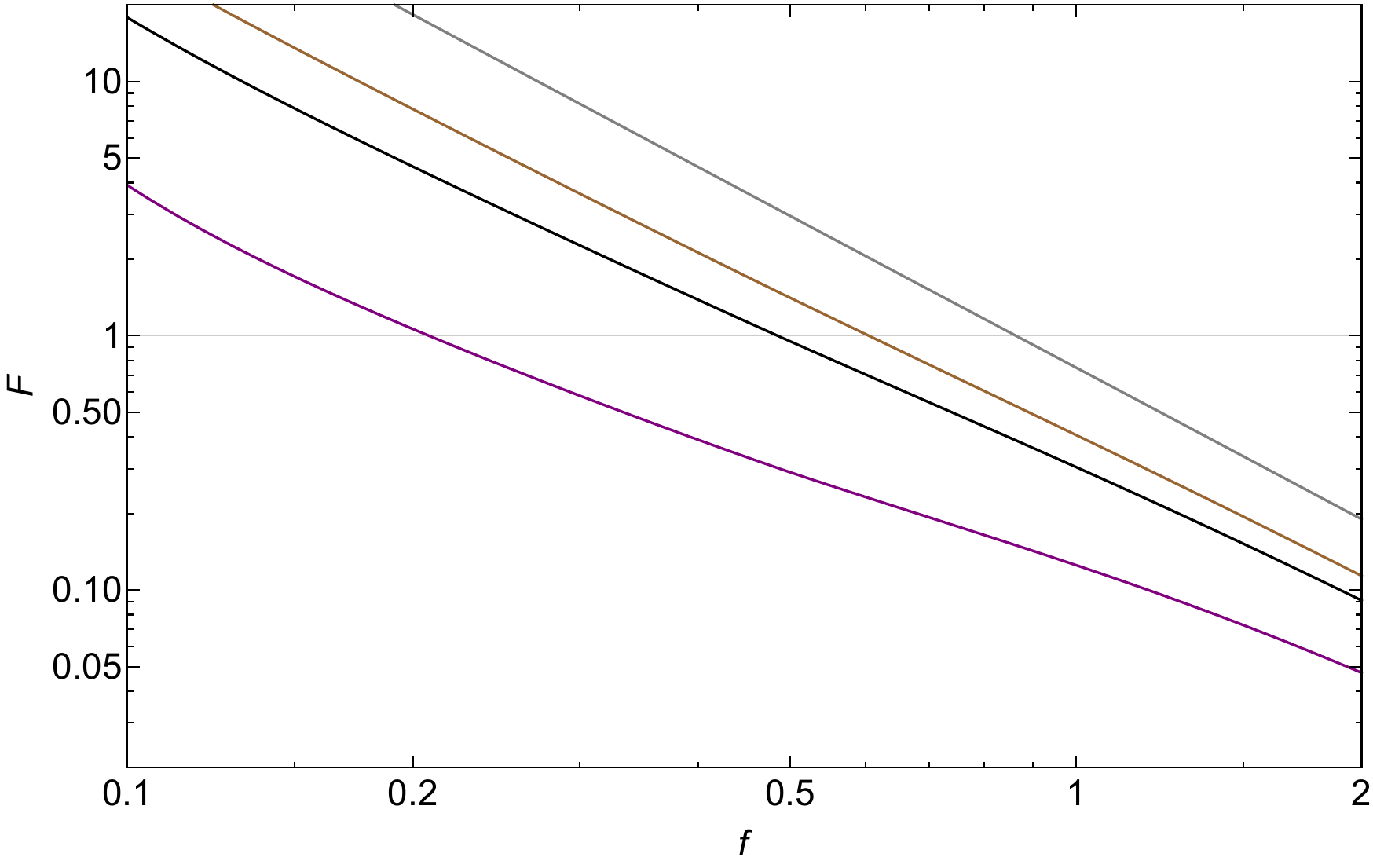}
    \caption{The dependence of the ratio of total mass in each species $F = M_\psi/M_\chi$ on the ratio of central field values $\rr = \Phi(0)/\Psi(0)$ for the groundstate solitons of the SP-system. We plot the curves for various mass ratios: $m_2/m_1 = 1/5$ (purple), $m_2/m_1 = 2/5$ (black), $m_2/m_1 = 1/2$ (brown) and $m_2/m_1 = 4/5$ (grey). As can be understood intuitively, the lighter field will tend to dominate the total mass sooner due to its large De Broglie wavelength. A gridline is added as a visual aid, when $F = 1$, or when both fields have the same total mass in the soliton solution.}
    \label{fig:Fvsf}
\end{figure}

In Fig.~\ref{fig:groundstatesolutionsm20d5} top panel and bottom panel, we show some ground state solitons with $m_2/m_1 = 1/2$ and $m_2/m_1 = 1/5$ respectively. 
We see that there is nontrivial behavior. In particular, the cyan curve crosses the magenta and orange curves for the second field $\Phi(r)$. We can understand this as follows: The cyan curve for $\Phi$ is so high (for small $r$) that it carries a large amount of mass. This means the fields are concentrated around this heavy center. However, as we lower the central value in $\Phi$ to the magenta or orange curve, the mass is reduced, the gravitational pull is reduced, and the fields are more spread out. In Fig.~\ref{fig:massdensityprofs} we plot the mass density profiles for a solution with a comparable total mass in the two fields ($O(50\%)$ difference). These correspond to the magenta (top panel) and orange (bottom panel) in Fig.~\ref{fig:groundstatesolutionsm20d5} respectively. The larger De Broglie wavelength of the light field is visible, resulting in an interesting density profile, that is initially dominated by the heavy field, and then gets dominated by the lighter field. Due to this wave effect, the lighter field can dominate the total mass of the system despite a small central density by concentrating particles farther away from the origin. This introduces an interesting question: how does the ratio of the total mass of the solitons, hereby referred to as $F = M_\psi/M_\chi$, depend on the ratio of central field values $\rr$? We plot this information in Fig.~\ref{fig:Fvsf} for various ratios of fundamental masses $m_2/m_1$. As expected, the smaller the ratio $m_2/m_1$, the sooner the lighter species tends to dominate the mass of the soliton. Interestingly, the dependence seems to follow an approximate power law.

Given that indeed these static solutions of the Schrödinger-Poisson system exist, we can investigate their dynamics. In particular, are the solutions that we found actually stable? Second, do they form dynamically under generic conditions? We address the first point in Appendix \ref{sec:appendixA}, and the second will be the main concern of the next section.

\section{Dynamical Evolution, Virialization \& Halo Formation}
\label{sec:Dynamical}
It is well known that single-field ULDM dynamically forms virialized Dark Matter halos with an approximately constant density core and an NFW-like tail. The core can be thought of as a highly populated ground state soliton, similar to those found in section \ref{sec:staticsols}. The ULDM halos that form in (cosmological) simulations have been argued to match observations of rotation curves better than standard CDM by suppressing small-scale structure. Some of the relevant issues are \textit{Core vs. Cusp, Missing Satellite}, and \textit{Too Big To Fail Problems}. It's important to check that multi-field ULDM still forms halos with these properties, as there would be no reason to consider them if they do not. 

In this section, we study the dynamics of multi-field ULDM enforcing spherical symmetry, by performing simulations of the virialization of a cloud of two-field ULDM. We acknowledge the limitations of enforcing spherical symmetry. However, as we'll show, our results are highly suggestive and provide enough evidence to make cosmological extrapolations.
\subsection{Numerical Experiments}
We are interested in answering two questions concerning the multi-field Schrödinger Poisson system of equations. Firstly, is the system able to virialize from generic initial conditions, forming a cored solitonic center with an NFW tail? Then if this does indeed happen, can we identify a relationship between the properties of the soliton core and the halo as a whole? In \cite{Schive_2014, Schive_20142} the relation between the core and halo masses in single-field ULDM was first discussed. It has since then been discussed in numerous works%reference
. We are interested in whether similar dependencies exist in the multi-field system.

To study these questions we perform numerical simulations of the Schrödinger Poisson system of Eqs.~\eqref{eq:SP1},\eqref{eq:SP2} and \eqref{eq:SP3} with an Runge-Kutta-4 integrator for time integration. At each time step, we solve the Poisson equation with and sixth-order accurate ODE solver. % using Eq.~\eqref{eq:dimensionlessphi1} and \eqref{eq:dimensionlessphi2} to turn the equations dimensionless. 
We use the following change of variables to obtain a dimensionless set of equations
\beq
\tilde{x}^{\mu} = m_1 x^\mu \quad \textrm{;} \quad \tilde{\psi} = \left(\frac{\sqrt{4 \pi}}{\sqrt{m_1} M_{pl}}\right) \psi\quad \textrm{;} \quad \tilde{\chi} = \left(\frac{\sqrt{4 \pi}}{\sqrt{m_1} M_{pl}}\right) \chi
\label{eq:pottrans2}
\eeq
Where tildes refer to dimensionless quantities. The SP-equations then become
\beq
i \dot{\tilde{\psi}} = - \frac{\nabla^2}{2} \tilde{\psi} + U \tilde{\psi}
\label{eq:SP1dimls}
\eeq
\beq
i \dot{\tilde{\chi}} = - \frac{m_1 \nabla^2}{2 m_2} \tilde{\chi} + \frac{m_2}{m_1} U \tilde{\chi}
\label{eq:SP2dimls}
\eeq
\beq
\nabla^2 U = \left(|\tilde{\psi}|^2 + \frac{m_2}{m_1} |\tilde{\chi}|^2\right)
\label{eq:SP3dimls}
\eeq
It is also noteworthy that under this change of variables we get natural definitions of dimensionless masses and energy related to dimensionful quantities as $\tilde{M} = \ \left(m_1/M_{pl}^2\right) M$ and $\tilde{E} = \left(m_1/M_{pl}^2\right)E$. Eqs.~\eqref{eq:SP1dimls}, \eqref{eq:SP2dimls} and \eqref{eq:SP3dimls} only explicitly depends on $m_1/m_2$. As long as the non-relativistic description of the scalars is valid, the overall dynamics thus only depend on the ratio of fundamental masses.
In what follows we drop the tilde and work with dimensionless quantities unless explicitly stated.
We investigate the virialization starting from two distinct types of initial conditions.
\begin{enumerate}
    \item Gaussian field profiles parameterized by 
    \beq
        \psi(r, 0) = \left(\frac{16 M_\psi^2}{\pi R_\psi^6}\right)^\frac{1}{4} e^{-\frac{1}{2}\left(\frac{r}{R_\psi}\right)^2}
        \label{eq:gaussianinitpsi}
    \eeq
    \beq
        \chi(r, 0) = \left(\frac{16 m_1^2 M_\chi^2}{\pi m_2^2 R_\chi^6}\right)^\frac{1}{4} e^{-\frac{1}{2}\left(\frac{r}{R_\chi}\right)^2}
        \label{eq:gaussianinitchi}
    \eeq
    Where $M_\psi$, $M_\chi$, $R_\psi$ and $R_\chi$ are free parameters. Note that the phase of the complex fields is 0 everywhere initially. The fields are thus identically real at $t = 0$.
    \item Random initialization of a gas of particles with specific velocity dispersion. Specifically, we initialize in Fourier space with,
    \beq
    \psi(|k|, 0) = \psi_{k, re} + \psi_{k, im} i
    \label{eq:randpsi}
    \eeq
    \beq
    \chi(|k|, 0) = \chi_{k, re} + \chi_{k, im} i
    \label{eq:randchi}
    \eeq
    Where the different coefficients of $\psi_k$ and $\chi_k$ are taken as Gaussian variables with variance given by $\sigma_\psi^2 = 2^\frac{7}{2} \pi^\frac{5}{2} M_\psi e^{-\frac{k^2}{2}}$ and $\sigma_\chi^2 = 2^\frac{7}{2} \pi^\frac{5}{2}\left(\frac{m_1}{m_2}\right)^4 M_\chi e^{-\frac{m_1^2 k^2}{2 m_2^2}}$. Once all the coefficients are determined we perform an inverse Fourier transform to obtain the initial conditions in real space. This initialization will more closely model the randomness to be expected of an over-density of Dark Matter before collapsing. However, we noted that they generally take longer to collapse, and the Gaussian profile is still valuable in terms of computational efficiency.
\end{enumerate}
For both type of initial conditions, we have another requirement on the parameters that we use to set the initial field profiles. Namely, we require the system to be bounded at the initial time; thus the total energy is smaller than $0$.
Finally, it is important to note that not necessarily all particles in our initial conditions will tend to collapse and produce a virialized halo. Some will tend to leave the gravitationally bounded domain towards infinity. As we only possess finite computational power, we accommodate this by defining a physical box in which we measure properties of the halo, surrounded by an absorbing boundary layer, in which escaped particles can decay. To be explicit, we follow the procedure outlined in \cite{Guzm_n_2004}.  
Having discussed our numerical setup, we now move on to our results. In sec.~\ref{sec:empericalfinds} we discuss our findings from the numerical time evolution of the two-field SP-system. First, in sec.~\ref{sec:emperics1}, we show that the two-field system, like its single field counterpart, generically forms a virialized halo with a groundstate solitonic core at its center. Then, in sec.~\ref{sec:halosoliton} we highlight a possible relation between halo and core that allows us to find the unique soliton we expect to be present at the center of a specific galaxy, given its conserved quantities, $E_h$, $M_\psi$ and $M_\chi$.

\subsection{Empirical Findings}
\label{sec:empericalfinds}
\begin{figure}
    \centering
    \includegraphics[width=\textwidth]{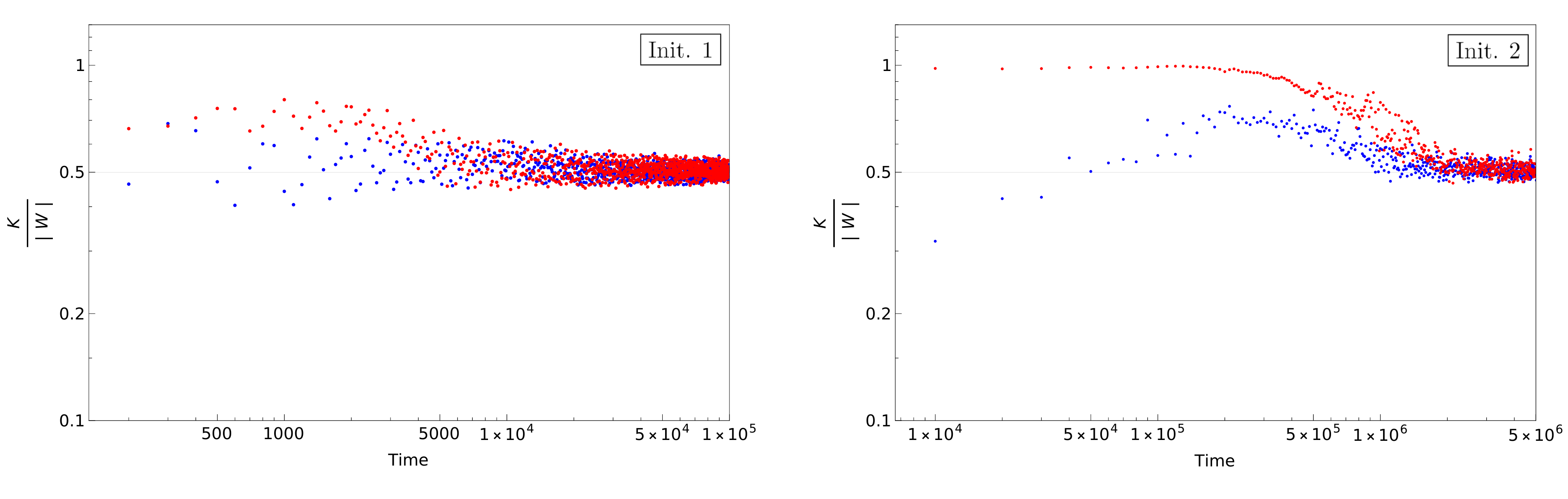}
    \caption{Typical evolution of the virial coefficient $K/|W|$ for undercooled (meaning $K/|W|<0.5$ initially) (blue) and overcooled (meaning $K/|W|>0.5$ initially) (red) initial conditions. Although the undercooled situation oscillates around $0.5$ earlier, eventually both situations tend to virialization, in a sense carrying no memory of the initial conditions. \textit{Left:} Initialization from a Gaussian packet following Eqs.~\eqref{eq:gaussianinitpsi} and \eqref{eq:gaussianinitchi}. \textit{Right:} Initialization from random initial conditions following Eqs.~\eqref{eq:randpsi} and \eqref{eq:randchi}. A Gaussian initialization takes about one order of magnitude less time to form a virialized configuration and hence offers large computational benefits.}
    \label{fig:virialcoefficient}
\end{figure}

The virial theorem states that $E_{kin} = \frac{1}{2} |E_{grav}|$ for a stable gravitationally bound system. In our setup, we are thus interested in configurations that obey
\beq
\frac{K}{|W|} = \frac{K_\psi + K_\chi}{|W_\psi + W_\chi|} = \frac{1}{2}
\label{eq:virialcond}
\eeq
within our numerical domain. It is not obvious that the virial theorem applies to ULDM halos. An elegant proof of this is given in \cite{Hui_2017}, which is easily generalizable to multi-field systems. In our simulations, we found that, regardless of initialization, the system virializes. We show the evolution of the virial coefficient $K/|W|$ in Fig.~\ref{fig:virialcoefficient}. In practice, we evolve from our initial conditions until sufficient time has passed for the system to virialize and become approximately static. We then take measurements of this ``model'' halo which has formed in our numerical domain.

\subsubsection{The formation of the halo}
\label{sec:emperics1}
\begin{figure}
    \centering
    \includegraphics[width = \textwidth]{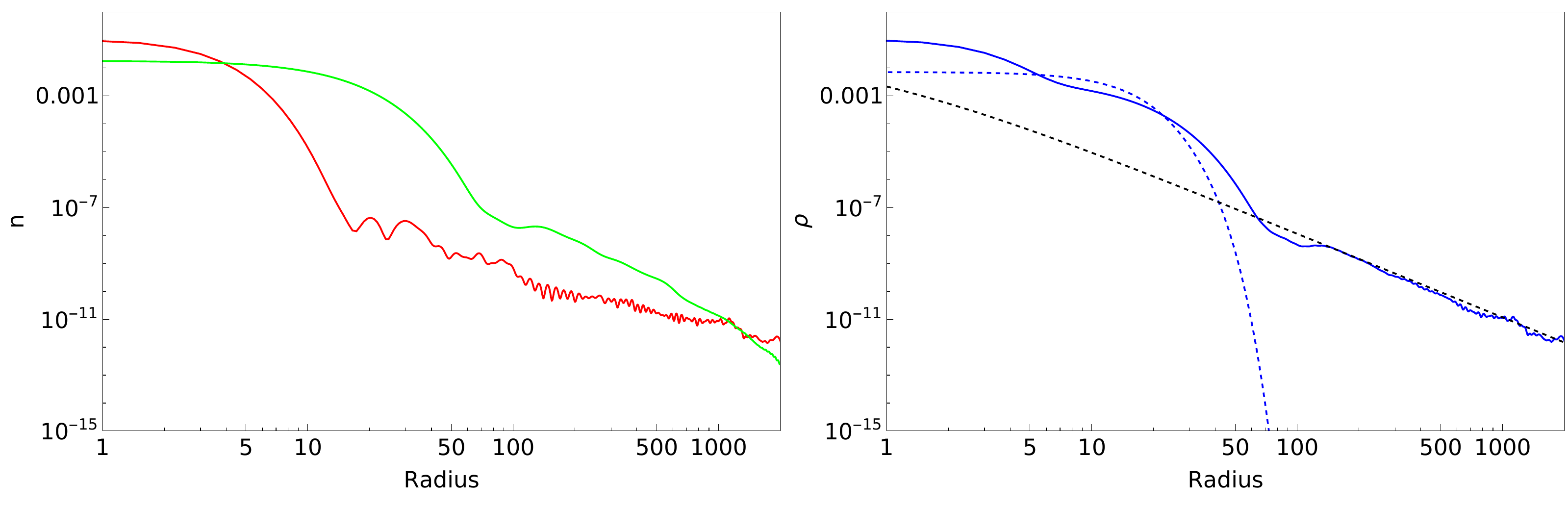}
    \caption{The (spherical) gravitational collapse of our two-field system, starting from a Gaussian packet following Eqs.~\eqref{eq:gaussianinitpsi} and \eqref{eq:gaussianinitchi} with $m_2/m_1 = 1/5$. \textit{Left:} The time-averaged number density profiles of the heavy (red) and light (green) fields. We see that the fields form a cored halo with a decaying tail. The decaying tail is not exactly NFW although it clearly follows some power law. \textit{Right:} The initial (dashed) and final time-averaged (full) mass density profiles. The tail of the full mass density profile follows NFW almost exactly (black dashed line). The nested ground state soliton is also clearly visible.}
    \label{fig:haloformGauss}
\end{figure}

\begin{figure}
    \centering
    \includegraphics[width = \textwidth]{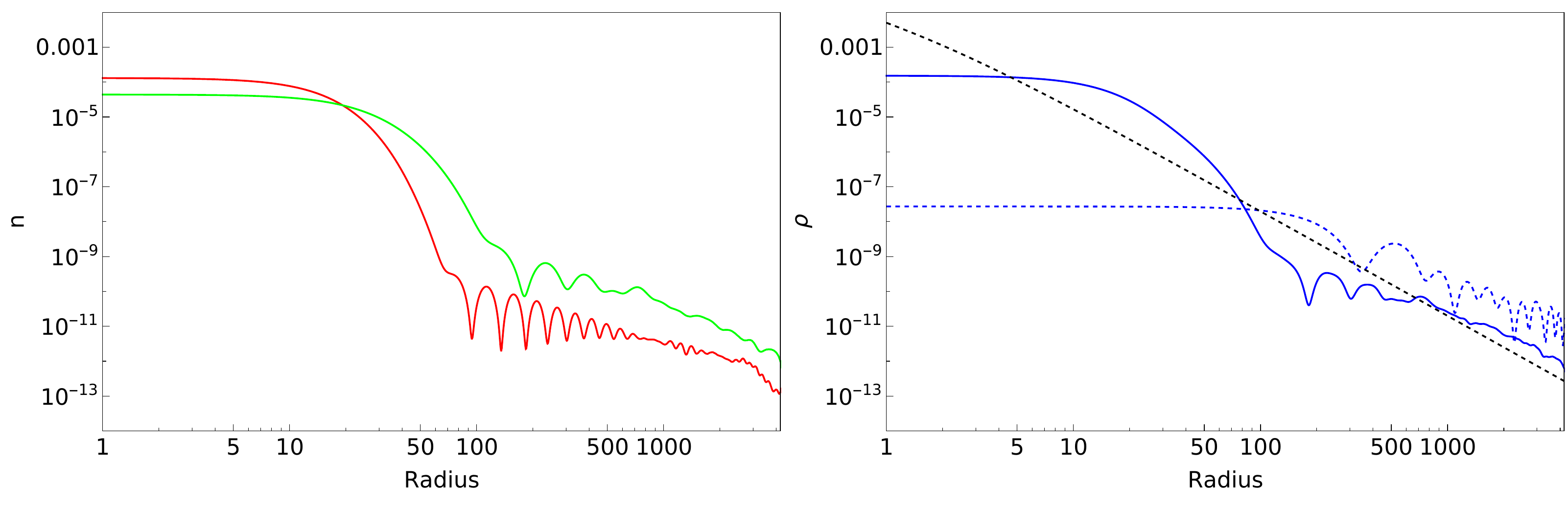}
    \caption{The same plots as in Fig.~\ref{fig:haloformGauss} with $m_2/m_1 = 1/2$ and starting from random initial conditions through Eqs.~\eqref{eq:randpsi} and \eqref{eq:randchi}. The initial conditions do not seem to greatly impact the final result of the simulation: a virialized cored halo with an NFW tail. Here we can also clearly see the analogy to density granules in spherical symmetry: dips and peaks in the profile supported by pressure gradients.}
    \label{fig:haloformrand}
\end{figure}

Regardless of initial conditions, eventually, a regime is reached where condition~\eqref{eq:virialcond} is obeyed. We show this for some typical initial conditions in Fig.~\ref{fig:virialcoefficient}. The remaining gravitationally bound structure can be seen as a model Dark Matter halo (or Boson star) supported by ``Quantum'' pressure. Virialization happens through a combination of gravitational collapse and the ejection of fast-moving matter to the absorbing boundary layer. In this work, we do not make statements about the precise timescales involved in these processes. However, we note that virialization proceeds more efficiently in the case where the initial conditions are ``undercooled'' ($k/|W| < 1/2$) as opposed to ``overcooled'' ($k/|W| > 1/2$). In the latter case, our setup tends to eject more matter into the absorbing layer, requiring more time. Understanding all the timescales involved in condensing into a two-field bound halo, is a very important endeavor, but beyond the scope of this work. We plan to address these questions in the future with more representative 3D simulations. 
In Figs.~\ref{fig:haloformGauss} and \ref{fig:haloformrand} we show the process of virialization for typical cases of each type of initialization highlighted in the previous section. For these simulations, we took two benchmark mass ratios of $m_2/m_1 = 1/2$ and $m_2/m_1 = 1/5$.
The halo has the properties that we expect from ULDM: a centralized core with an NFW tail. Assuming that the typical velocities of the particles in the halo are the same, the lighter field has a larger De Broglie wavelength $\lambda = 1/mv$. This can be seen especially in the core where the lighter field has a broader density profile. The overall mass density can thus transition between being dominated by the more massive field to being dominated by the lighter field, potentially leading to interesting observational signatures. Lastly, note that even though virialization happens faster in the case where we initialize with a Gaussian field profile (which makes intuitive sense as the phases of the fields are already correlated over large distances initially), the virialized final products of the simulations have similar properties across different initializations, in a way ``erasing'' the memory of the initial conditions.

\begin{figure}
    \centering
    \includegraphics[width = \textwidth]{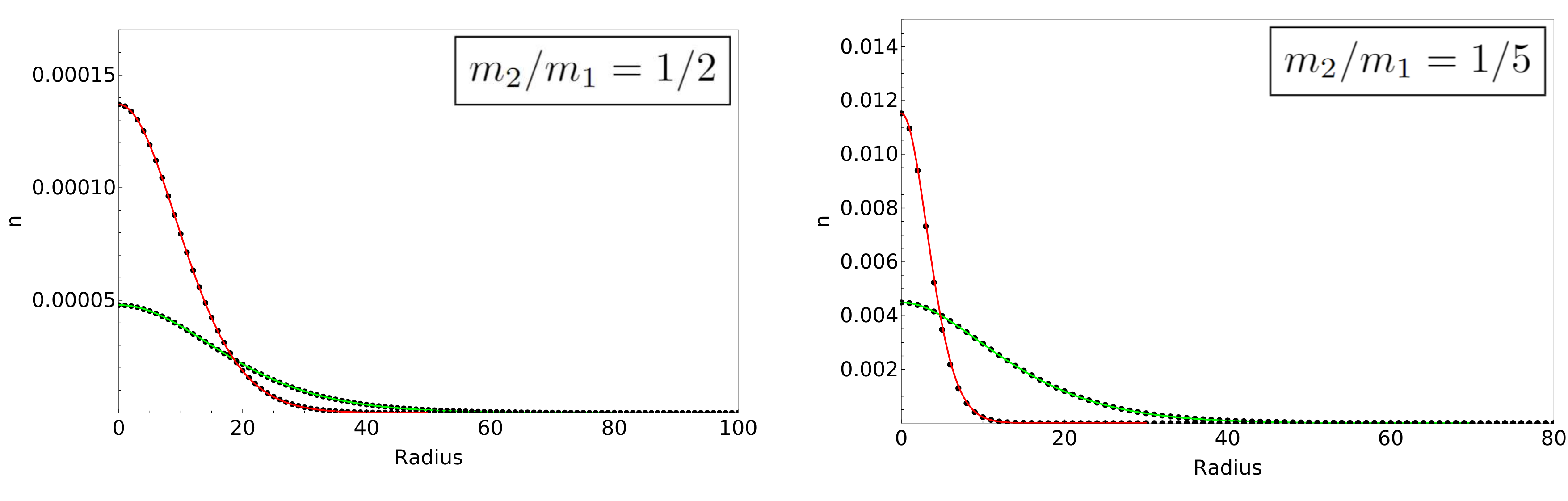}
    \caption{The number density profile of the heavy (red) and light (green) fields at the end of two of our simulations. In black dotted, we plot the corresponding two-field soliton that matches the central number density of the two fields. The core of our halo is exactly one of the gravitational solitons found earlier in this work. Although here we only show the product of two simulations, we found that a nested two-field soliton can always be found at the center of a properly virialized halo, including for different mass ratios $m_2/m_1$. \textit{Left:} $m_2/m_1 = 1/2$ \textit{Right:} $m_2/m_1 = 1/5$.}
    \label{fig:gravitsolform}
\end{figure}

An important question remains: are the cores of these halos the gravitational solitons found in Sec.~\ref{sec:staticsols}? To check this we take the final ratio of the density profiles of the two fields and use the algorithm of \ref{sec:staticsols} to solve for the corresponding ground state solitons. The results are shown in Fig.~\ref{fig:gravitsolform}
It seems clear that the core of the halo can indeed be interpreted as a static solution of the two-field Schrödinger-Poisson system. Although these solitons are in principle static when isolated, we observe oscillation around the true static solution of $O(10\%)$, due to interactions with the halo. This is not altogether unexpected and can be explained as a wave-interference effect as was done in \cite{Li_2021} and \cite{Lin_2018}. We also want to note that Refs.~\cite{Huang_2023, luu2023nested} reported on certain situations (depending on the relative abundance of particles and the ratio of fundamental masses $m_2/m_1$) where the two-field soliton was not able to form due to tidal interactions between the different components during the formation of the halo. We wish to report that we did not observe this in our simulations. In particular, we always observed a two-field soliton at the center of the virialized halos that we generated. However, we acknowledge that this might in part be due to the limitations of imposing spherical symmetry on our system.

Although we now have seen that a two-field soliton forms at the center of virialized halos, we still have no way to know what particular soliton should be present in what halo (e.g. what is the ratio of central densities of the two fields). In other words, what is the thermodynamic equilibrium of the halo-soliton system. To constrain multi-field models of ULDM, this is an important question to answer. Namely, we are interested in relating the properties of the halo to the solitonic core.

\subsubsection{The relationship between the halo and the soliton}
\label{sec:halosoliton}
In studies of structure formation of (single-field) ULDM models, an interesting relation was discovered between core mass $M_c$ (mass contained in the region where the mass density remains $50\%$ of the central density and the halo mass $M_h$. Previous work \cite{Schive_20142, Schive_2014} found the following scaling 
\beq
M_c \propto a^{-1/2} M_h^{1/3}
\label{eq:corehalomassrel}
\eeq
where $a$ is the scale factor of the Universe. This relation has been somewhat disputed in the literature \cite{Mocz_2017, Bar_2018, Levkov_2018, dmitriev2023selfsimilar, dmitriev2024selfsimilar}, but most groups agree that some type of scaling is present, with \eqref{eq:corehalomassrel} the most commonly cited one. Somehow, the wave nature of ULDM connects the properties of the central soliton to those of the enveloping halo. The question has to be asked whether a similar connection holds when considering multiple fields. In \cite{Bar_2018} it was suggested that the scaling \eqref{eq:corehalomassrel} can be explained through an empirical ``thermodynamic'' relation that is obeyed in the halo, namely
\beq
\frac{|E_h|}{M_h} = \frac{|E_s|}{M_s}
\label{eq:thermodynamicssolhal}
\eeq
The energy per unit of mass has the same value in the halo and the central soliton. Using \eqref{eq:thermodynamicssolhal} together with some considerations from collapse models of overdensities, one can arrive at \eqref{eq:corehalomassrel} rather straightforwardly. However, this is in large part due to the simple scaling of the single-field solitons, and a simple scaling like \eqref{eq:corehalomassrel} is not expected to exist in our two-field system. 

\begin{figure}[t!]
    \centering
    \includegraphics[width = \textwidth]{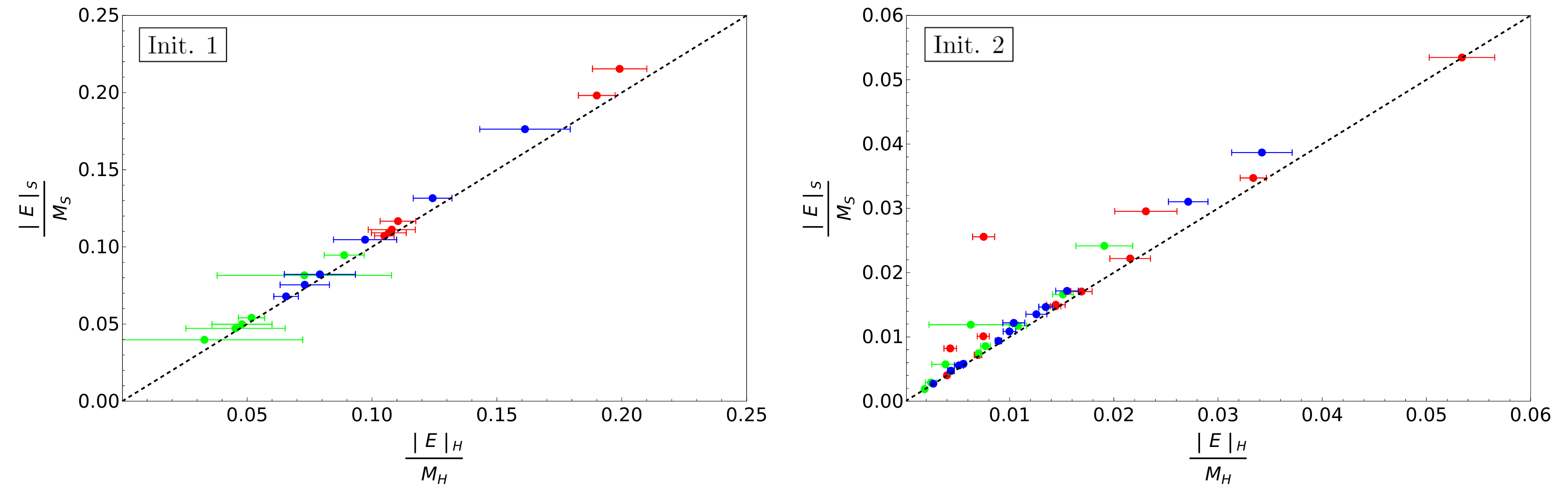}
    \caption{The energy per unit mass for the full halo obtained as the final product of our simulation (y-axis) versus the energy per unit mass for the soliton at the center (x-axis), for $m_2/m_1 = 1/2$. We plot the values for the heavy field (red), light field (green), and the full system (blue). In black dotted, we plot the line $y = x$. Although there remains large uncertainty, these results are highly suggestive. \textit{Left:} starting from a Gaussian packet following Eqs.~\eqref{eq:gaussianinitpsi} and \eqref{eq:gaussianinitchi}. \textit{Right:} The same plot but for the halos obtained from random initialization following Eqs.~\eqref{eq:randpsi} and \eqref{eq:randchi}.}
    \label{fig:gaussianinitEpM}
\end{figure}

\begin{figure}[t!]
    \centering
    \includegraphics[width = \textwidth]{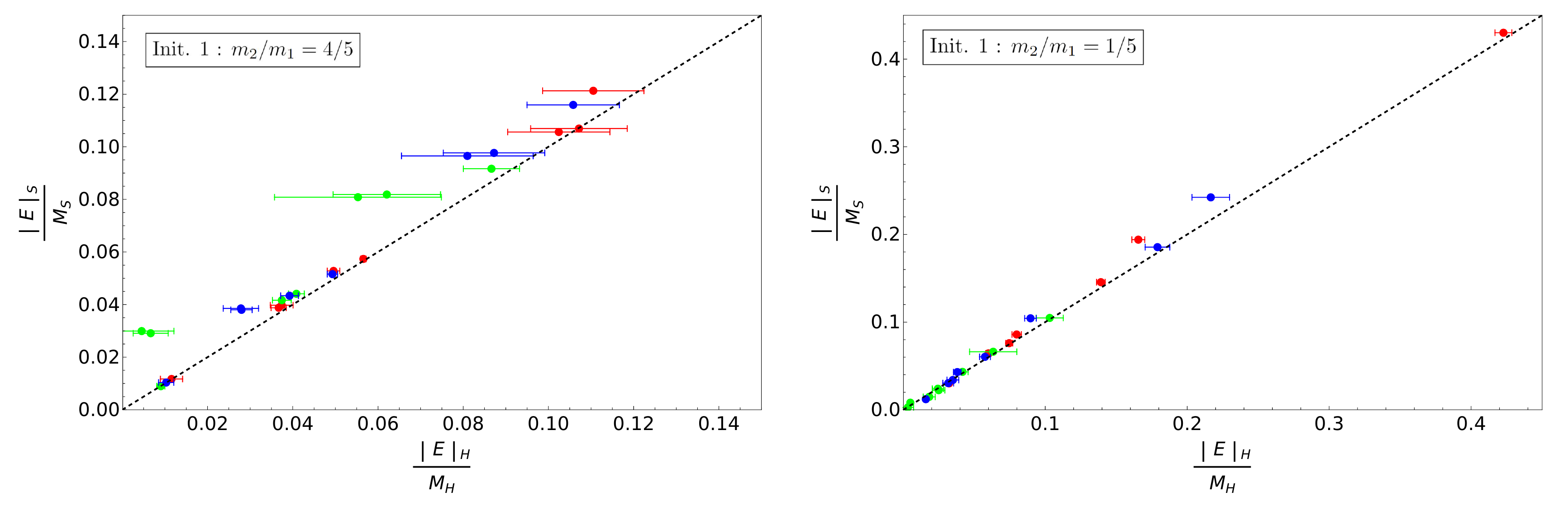}
    \caption{The same data as plotted in Fig.~\ref{fig:gaussianinitEpM} starting from a Gaussian packet following Eqs.~\eqref{eq:gaussianinitpsi} and \eqref{eq:gaussianinitchi} for different mass ratios. We choose the Gaussian initialization because it tends to relax to a virialized halo faster, and thus speeds up computations. \textit{Left:} $m_2/m_1 = 4/5$. \textit{Right:} $m_2/m_1 = 1/5$. Qualitatively, the results agree with Fig.~\ref{fig:gaussianinitEpM}, namely that the energy per unit mass has the same value in the soliton, as it has in the halo.}
    \label{fig:randominitEpM}
\end{figure}

However, we can ask if a relation like \eqref{eq:thermodynamicssolhal} still exists. In the multi-field case, there are more potential relations to be explored. In particular, we can consider the energy per unit mass in the system as a whole, but also at the level of each field individually. We checked this for the different virialized halos of our simulations by comparing the properties of the halo in the full numerical box with the ones of the soliton core. The properties of the core are taken by solving for the appropriate soliton solution, using the algorithm of Sec.~\ref{sec:staticsols}. We plot the results in Figs.~\ref{fig:gaussianinitEpM} and \ref{fig:randominitEpM}, separating the different types of initializations and mass ratios $m_2/m_1$ we have considered. 

Looking at Figs.~\ref{fig:gaussianinitEpM} and \ref{fig:randominitEpM}, there seems to be fairly compelling evidence for the following three empirical relations:
\beq
\frac{|E_{h, tot}|}{M_{h, tot}} = \frac{|E_{s, tot}|}{M_{s, tot}}; \quad \frac{|E_{h, \psi}|}{M_{h, \psi}} = \frac{|E_{s, \psi}|}{M_{s, \psi}}; \quad\frac{|E_{h, \chi}|}{M_{h, \chi}} = \frac{|E_{s, \chi}|}{M_{s, \chi}}
\label{eq:thermorel}
\eeq
Where the subscripts refer to the halo and soliton of the total system and the two fields individually. We want to place a caveat here, as we have to note that the radial simulations generally yielded halos whose masses were dominated by the central soliton. The relations in Eqs.~\eqref{eq:thermorel} are then satisfied somewhat trivially. Checking whether these relations emerge in general is left for the future as it requires 3D simulations that go beyond the scope of this work.

Interestingly, if the relations in \eqref{eq:thermorel} hold generically, one is completely able to determine which solitonic core is present in which galaxy, based on the conserved quantities of the system, namely $E_h$, $M_\psi$ and $M_\chi$. There is exactly one soliton solution that can satisfy these properties and the system is neither over or under determined. Let's outline an algorithm that can solve for a particular soliton solution, starting from $E_h$, $M_\psi$ and $M_\chi$.
\begin{enumerate}
    \item Using the algorithm of Sec.~\ref{sec:staticsols}, we obtain a function of $|E_{s,fr, tot}|/M_{s, f, tot}$ for the soliton solutions with central densities $\Psi_0 = 1$ and $\Phi_0 = f$. Through the scaling relations of Eqs.~\eqref{eq:masstrans}, \eqref{eq:kintrans} and \eqref{eq:pottrans}, we then know what the scaling parameter $\lambda$ should be for each ratio of central densities in order to satisfy $|E_{h, tot}|/M_{h, tot} = |E_{s, f, tot}|/M_{s, f, tot}$.
    \item Similarly, we can create functions $|E_{s, f, \psi}|/M_{s, f, \psi}$ and $|E_{s, f, \chi}|/M_{s, f, \chi}$. At each ratio of central density, we have to satisfy
    \beq
    |E_\psi| = \lambda^2 M_\psi \frac{|E_{s, f, \psi}|}{M_{s, f, \psi}} \quad \textrm{and} \quad  |E_\chi| = \lambda^2 M_\chi \frac{|E_{s, f, \chi}|}{M_{s, f, \chi}}
    \label{eq:constraintsolitons}
    \eeq
    Where at each ratio we have determined $\lambda$ in step 1.
    \item Realising that $E_\psi + E_\chi = E_h$ and using Eq.~\eqref{eq:constraintsolitons} we can rewrite Eqs.~\eqref{eq:constraintsolitons} to obtain 
    \beq
    |E_h| = \lambda^2\left( M_\psi \frac{|E_{s, f, \psi}|}{M_{s, f, \psi}} +M_\chi \frac{|E_{s, f, \chi}|}{M_{s, f, \psi}}\right)
    \label{eq:constraintsol}
    \eeq
    This equation will only be satisfied for one particular ratio of central densities (and through step 1 $\lambda$). We will then have found the appropriate soliton solution.
\end{enumerate}
This type of computation allows us to know what multi-field soliton should exist in a given halo when we know its conserved quantities. We will use this in the next section to discuss the cosmological implications of two-field ULDM.

\section{Cosmological Implications}
\label{sec:cosm}
\begin{figure}[t]
    \centering
    \includegraphics[height=8.5cm,width=9.5cm]{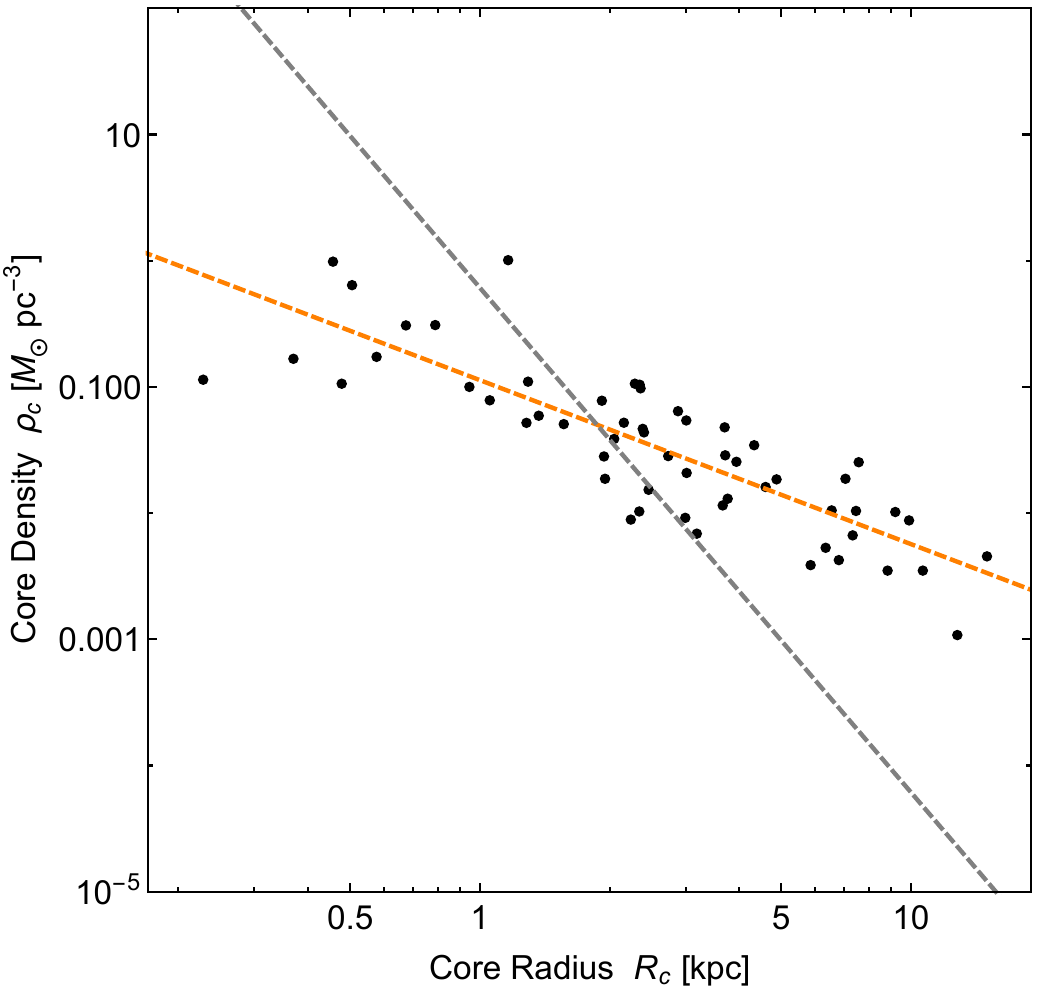}
    \caption{The best-fit parameters of the Core Density $\rho_c$ and Core Radius $R_c$ of 56 galaxies obtained from \cite{Rodrigues_2014}. The data suggests a scaling of $\rho_c \propto 1/R_c^\beta$ with $\beta\approx 1.3$ (see dashed orange line). This puts a strain on all single-field ULDM models, since they predict a value of $\beta$ significantly higher, namely $\beta=4$ (see gray dashed line). This apparent contradiction was first highlighted in \cite{Deng_2018} (and this figure is reproduced from Ref.~\cite{Deng_2018}, plus the added gray line). As we'll see, multi-field models alter this precise theoretical prediction.}
    \label{fig:obspcvsRc}
\end{figure}

\begin{figure}
    \centering
    \includegraphics[width = \textwidth]{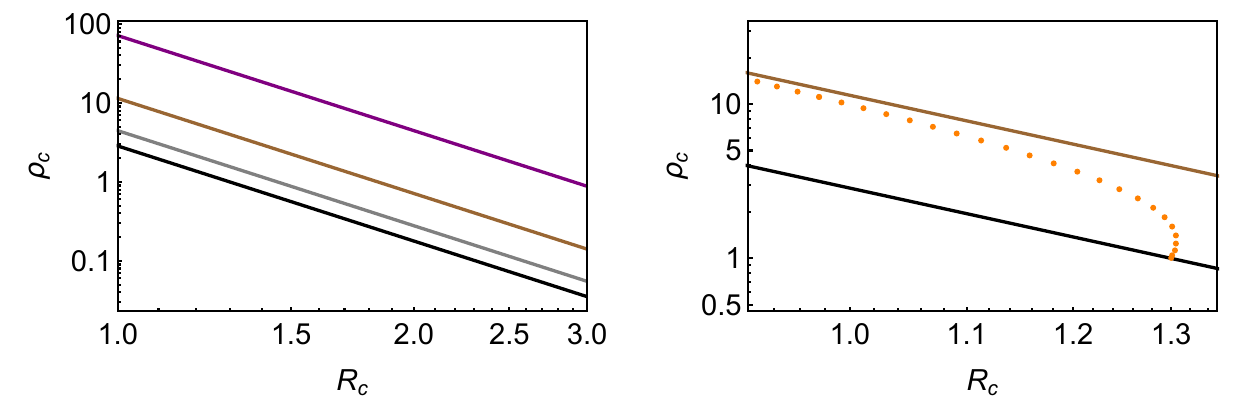}
    \caption{\textit{Left:} The various scalings of $\rho_c$ versus $R_c$ for single-field gravitational solitons in terms of dimensionless variables. The single-field solitons all follow a scaling of $\rho_c \propto 1/R_c^4$. Here we plot the scalings corresponding to $m_2/m_1 = 1$ (black, the line corresponding to the heavy field), $m_2/m_1 = 4/5$ (gray), $m_2/m_1 = 1/2$ (brown) and $m_2/m_1 = 1/5$ (purple). \textit{Right:} The full two-field gravitational soliton solutions with $m_2/m_1 = 0.5$ (orange scatter), where the ratio of central field values $\Phi_0/\Psi_0$ is increased from $0$ to $\infty$. The full solution interpolates between the two asymptotes thus opening up the entire region of parameter space enclosed by these lines.}
    \label{fig:greyregion}
\end{figure}

The major success of the ULDM paradigm lies in its ability to suppress structure on small scales, matching observations \cite{Rodrigues_2017} %more references
. The presence of a solitonic core at the center of collapsed halos is a strong prediction and its properties can in principle be used to obtain observational constraints. In fact, in previous work one of us argued that the properties of the single-field solitons disfavor most ULDM models \cite{Deng_2018}. Central to the argument is the relation between the core central density $\rho_c$ and the core radius $R_c$ (the radius at which the soliton density drops by $50\%$). The observational results, when fitting rotation curves with a cored density profile are given in Fig.~\ref{fig:obspcvsRc}.
Fig.~\ref{fig:obspcvsRc} suggests a scaling of $\rho_c \propto 1/R_c^\beta$, with the best fit provided by $\beta \approx 1.3$. Now the problem arises: if the cores are provided by gravitational solitons of the ULDM field, there are no single-field models that contain such a scaling. In particular, if the scalar has no self-interactions, the scaling is always $\rho_c \propto 1/R_c^4$. This can be easily seen through the transformation of Eq.~\eqref{eq:psitrans}. Interestingly, one finds more promise in the multi-field models under consideration here, as was first pointed out in \cite{Guo_2021}. As there are distinct soliton solutions at each ratio of the central density (not related through a simple scaling relation), the solitons interpolate between the two asymptotes where one of the fields dominates the mass, and we are again in the regime of $\rho_c \propto 1/R_c^4$. In this way a whole new region of parameter space becomes available. We show this schematically in Fig.~\ref{fig:greyregion}.

We now want to see whether the two-field scenario can actually solve the issue, considering the previous sections of this work. In particular, do we expect the solitons that form, to actually have the correct scaling? We want to note that in a similar way as schematically outlined above, multi-component ULDM can ``escape'' other constraints that exist for single-field models. In particular in \cite{Bar_2018} a tension is highlighted, coming from rotation curve data. By opening up the parameter space of possible solitons that are present in the core, this apparent problem can also be alleviated.

To make comparisons with Cosmology we need to put back dimensions into our story. To do this we consider $m_1 = \SI{5e-22}{\eV}$. Different masses might be considered (and even match the data of Fig.~\ref{fig:obspcvsRc} better), but won't change the overall conclusions. As first noted in \cite{Guo_2021}, to accommodate all the galaxies in Fig.~\ref{fig:obspcvsRc}, we need at least $m_2/m_1 \approx 10^{-3}$. Since it is numerically intense to generate data about the core solitons for these values of the mass ratio, we limit ourselves to some benchmark cases. The overall, qualitative conclusions are valid for any value of the mass ratio\footnote{It is still an open question whether any two-field soliton can form realistically, which might put tension on the conclusions presented here; see \cite{Huang_2023, luu2023nested}.}.

\begin{figure}
    \centering
    \includegraphics[width = \textwidth]{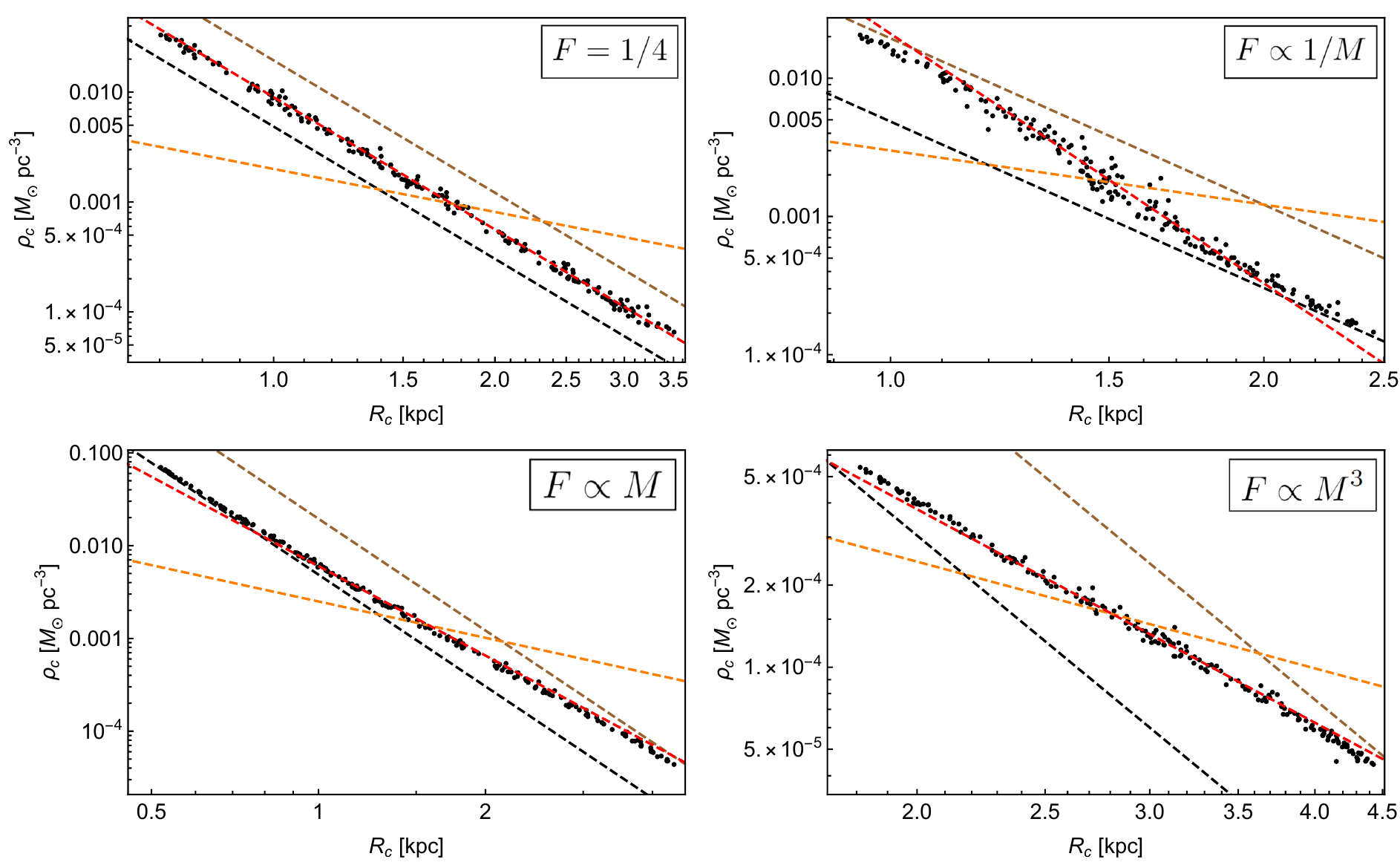}
    \caption{The properties of the solitons of our mock galaxy set as generated through the method highlighted in the text (black scatter). The properties of the single-field solitons of the heavy and light species are also shown (black and brown dotted). In each plot, we chose a different dependence for the ratio of total mass of each species $F$, with $F = 1/4$ (top left), $F \propto M$ (bottom left), $F \propto 1/M$ (top right) and $F \propto M^3$ (bottom right). The emperical scaling of $\rho_c \propto 1/R_c^{1.3}$ is shown as the orange dotted line. As expected, the species that dominates the mass of the galaxy also dominates the soliton properties. We can thus find solitons living in the region of parameter space between the two extremal asymptotes. The scaling of $\rho_c$ versus $R_c$ can then differ from the single-field story. The red dotted line gives the approximate new scaling that emerges and in particular, we have $\rho_c \propto 1/R_c^4$ (top left), $\rho_c \propto 1/R_c^{3.2}$ (bottom left), $\rho_c \propto 1/R_c^6$ (top right) and $\rho_c \propto 1/R_c^{2.6}$ (bottom right).  For the case that $F$ is positively correlated with the mass of the galaxy $M$, we see that the tension with Fig.~\ref{fig:obspcvsRc} can be alleviated.}
    \label{fig:scalingscosmo}
\end{figure}

Using the relations of Eq.~\eqref{eq:thermorel} and the algorithm discussed at the end of Sec.~\ref{sec:halosoliton} we can generate a mock galaxy data set containing the information about the soliton expected to be present at the center of the halo. To do this we need to input the conserved quantities: the total energy $|E_h|$ and the masses of the two fields $M_\psi$ and $M_\chi$. We will assume that the energy of the halo is that of a non-moving overdensity of particles (so no kinetic energy), with a constant overdensity profile given by $\rho = 10^{-5} \rho_{0}$ where $\rho_{0} =  \SI{1.8e-9}{\eV}^4$ matches the intergalactic (IGM) density of about 1 proton per cubic meter. The energy of such a cloud is then given by $E_h = -\frac{3}{5} \frac{G M^{5/3} \rho^{1/3}}{(3/4\pi)^{1/3}}$, with $M = M_\psi + M_\chi$ the total mass. To generate our set we then only need to make an assumption for the total mass in the galaxy $M$ and the relative abundance of the two species of particles $F = M_\psi/M_\chi$. To generate a mock sample of galaxies, we choose $M = 10^k M_{MW}$, where k is taken from a uniform distribution with boundaries between $[-1, 1]$. We use $M_{MW} = 10^{12} M_\odot$, the Milky Way mass. Finally, we need to choose $F$. The most natural thing might be to assume that the relative abundance does not change with the galactic mass. However, it might also be interesting to investigate the possibility of $F$ depending on $M$. In particular, we will look at $F \propto M$ and $F \propto 1/M$, as well as $F = \mathrm{Const}$. To investigate an extremely exotic scenario we also looked at $F \propto M^3$. The proportionality factor is an $O(1)$ number in all cases. We plot the expected solitons for the four scenarios in Fig.~\ref{fig:scalingscosmo}. In each case, we choose the value of $F$ from a Gaussian distribution centered around its expected average value $\mu_F$ ($\mu_F = 1/4$, $\mu_F \propto M$, $\mu_F \propto 1/M$, $\mu_F \propto M^3$, respectively) with a variance of $\sigma_F = 0.25 \mu_F$. \\

\begin{figure}
    \centering
    \includegraphics[width = \textwidth]{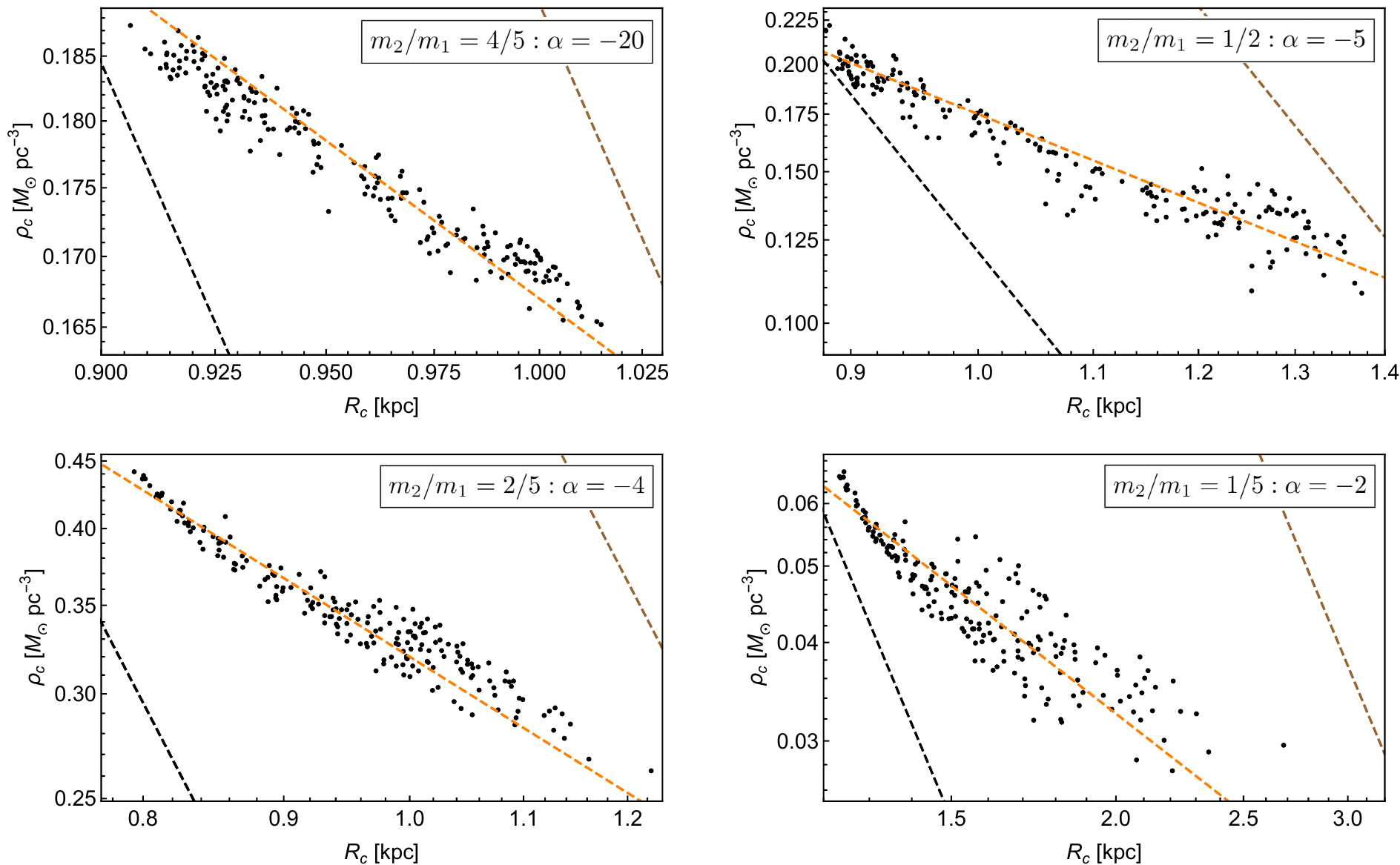}
    \caption{A similar plot as Fig.~\ref{fig:scalingscosmo}, but with $\alpha$ chosen using Eq.~\eqref{eq:beta} such that the correct scaling of $\beta = 1.3$ is obtained (orange dotted line). We plot this for various $m_2/m_1$: $4/5$ (top left), $1/2$ (top right), $2/5$ (bottom left), $1/5$ (bottom right). The values of $\alpha$ as (approximately) given by Eq.~\eqref{eq:beta} are $-20$, $-5$, $-4$ and $-2$ respectively.}
    \label{fig:correctscalings}
\end{figure}

Inspecting Fig.~\ref{fig:scalingscosmo} we see that non-trivial scalings only emerge when the ratio of masses in the galaxies is dependent on the total galaxy mass. This makes intuitive sense: if the ratio of masses is the same across every halo, one would expect that the ratio of core densities of the two fields is also the same. This is what happens in our mock data set. In this case, we are back in the scaling regime through the transformations of Eqs.~\eqref{eq:psitrans}, \eqref{eq:phitrans} and \eqref{eq:newttrans} and the soliton central density scales with $\rho_c \propto 1/R_c^4$. 
To reconcile the two-field model with Fig.~\ref{fig:obspcvsRc} we need a non-trivial function for the mass ratio $F$. Allowing the halo to be dominated by different species in different limits of its total mass, causes the soliton solution to interpolate between the two asymptotic single-field solutions. This is the source of the non-trivial scaling. Of course, we could have chosen different functions for the ratio $F$, which would change the exact scaling of the galaxies in the interpolating region. It is important to understand what type of scaling one actually needs to obtain $\beta = 1.3$.\\It is possible to get an estimate for the scaling that emerges, given the dependence of $F\propto M^\alpha$ and the ratio of fundamental masses $m_2/m_1$. In appendix~\ref{sec:appendixB} we show how to get the following approximate formula for $\beta$
\beq
\beta \approx -\frac{\log(0.075^{4/3\alpha}) + \log(\left(\frac{m_2}{m_1}\right)^{2})}{\log(0.075^{-1/3\alpha}) + \log(\left(\frac{m_2}{m_1}\right)^{-1})}
\label{eq:beta}
\eeq
Which we note has quite a considerable amount of uncertainty due to the exponential sensitivity of the parameters (see appendix ~\ref{sec:appendixB}). Eq.~\ref{eq:beta} has some interesting properties. Besides predicting the approximate scaling parameter $\beta$ to be expected at any mass ratio $m_2/m_1$ and $\alpha$, it also contains information about the ratio of the total mass of the heaviest and lightest galaxies. Furthermore, it also implies a correlation between the total mass of a galaxy and its core properties. These aspects are discussed in more detail in appendices~\ref{sec:appendixB} and \ref{sec:appendixC}. Using Eq.~\eqref{eq:beta} we can find the necessary $\alpha$ to reproduce $\beta = 1.3$ for any $m_2/m_1$. Using this we can create mock galaxy sets that obey the correct scaling of $\beta = 1.3$. We show this in Fig.~\ref{fig:correctscalings}. The necessary value of $\alpha$ is predicted by Eq.~\eqref{eq:beta}. Unfortunately, it is numerically difficult to check whether Eq.~\eqref{eq:beta} holds for even smaller ratios of the masses $m_2/m_1$ but in what follows we can speculate taking into account these large uncertainties. 
A particularly interesting case has (as first pointed out in \cite{Guo_2021}) $m_2/m_1 = 10^{-3}$ since for $m_1 \sim \SI{e-21}{\eV}$ and $m_2 \sim \SI{e-24}{\eV}$ all the datapoints in Fig.~\ref{fig:obspcvsRc} can be fitted by a groundstate soliton. Essentially, all the data falls in between the two extremal single-field lines. For this mass ratio, we find that $\alpha \approx -0.5$ reproduces the correct scaling. It would thus be necessary for more massive galaxies to contain a larger abundance of the lighter species of particles. It is clear, that some nontrivial scaling of $F\propto M^\alpha$ is needed. Interestingly, for $m_2/m_1 \sim 10^{-3}$, our model predicts that the ratio of masses of the heaviest and lightest galaxies be about $\sim O(10^2-10^3)$, matching good estimates of the data of the galaxies considered in Fig.~\ref{fig:obspcvsRc} (also see appendix~\ref{sec:appendixC}).
In the conclusions, we comment on whether scenarios, where the relative abundance of particle species varies across different galaxies, can be accommodated in viable cosmological models. A detailed study structure formation in multi-wave ULDM goes beyond the scope of this work, although we plan to address it in future studies. It is clear, however, that multi-field models do not trivially solve the scaling issue highlighted in \cite{Deng_2018}. On a more basic level the above exercise hints at other correlations to be tested against empirical data. In particular, the value taken for the halo energy $E_h = -\frac{3}{5} \frac{G M^{5/3} \rho^{1/3}}{(3/4\pi)^{1/3}}$, common in the literature, combined with Eqs.~\eqref{eq:thermorel} predicts the correlation between galaxy mass and central density of the core; and the correlation between galaxy mass and core radius. This is true for the multi-field models under consideration here but is equally valid when the core is comprised of just a single field. These correlations can thus be used to distinguish between the two. In appendix~\ref{sec:appendixC} we test this condition against the experimental data and highlight a tension that seems generic for ULDM models but can be alleviated by adding more fields to the picture. It is even plausible that one can consistently address both the scaling tension of Fig.~\ref{fig:obspcvsRc} and this other tension. We will now summarize and conclude this work.

\section{Conclusions and Outlook}
\label{sec:Conclusion}

In this work, we discussed many of the phenomenological features of the minimal extension of the standard Ultra Light Dark Matter paradigm with an additional light scalar. Since our results only depend on the ratio of fundamental masses $m_2/m_1$ we mostly focused on a dimensionless analysis in which we wanted to paint the imprints of this model in broad strokes. We plan to perform a more comprehensive study of large scale structure formation in the future. The first order of business for any extension of ULDM is to see whether it is successful in suppressing structure on small scales since this is the motivation for the original model. Through a large number of simulations, albeit performed in spherical symmetry,  we were able to confirm that this is indeed still the case for the two-field model under consideration. Starting from a variety of initial conditions, we found that virialized structures form eventually, with a centralized core matched to an NFW-like tail that extends to the edge of the halo. The core, a characteristic feature of ULDM can be identified with the groundstate soliton of the now multi-field SP-system. To confirm this we developed an iterative procedure to find eigenstates, referred to as gravitational solitons. Although we focused primarily on the lowest energy state, the method can easily be extended to find excited states, and even to other models which include (self-)interactions. Interestingly, these cores typically show a double feature in their density profile, where the heavier field dominates the soliton near the galactic center, but then gets dominated by the lighter species at some crossover point due to the larger De Broglie wavelength of the latter. To the best of our knowledge, this type of feature has not yet been explored (extensively) in the fitting of galactic rotation curves. Since this type of feature seems to be a generic prediction of multi-field wave Dark Matter, this type of exercise would be an interesting avenue for future research.

In previous work \cite{Deng_2018}, it was demonstrated that single scalar field models do not fit the core-cusp data in galaxies; the problem is that although one can fit a single galaxy by the appropriate choice of particle mass $m$, the trend among galaxies is incorrect. In particular, the core density versus core radius relation is observed to be $\rho_c\propto 1/R_c^\beta$, with $\beta\approx1.3$, however, the single field model predicts $\beta=4$.
Therefore it is important to consider the multi-field case, as done here. In this case, the ratio of the fundamental particle masses provides a parameter in the problem and so does the ratio of total mass in the galaxy. Relatedly, there is a much larger space of soliton solutions, compared to the single field case.

In \cite{Guo_2021} it was first hinted that the issue might be addressed in a multi-field model. However, to show this, one needs to understand which core solution is expected to be present in any particular halo. In other words, what is the thermodynamic equilibrium state of the halo-soliton system? This question is yet to be definitively answered in the single-field case and is probably even more complex in our model. In our simulations, we find good agreement with the emperical relations of Eqs.~\eqref{eq:thermorel}, which we use to develop an algorithm to predict the core solution of any halo based on its conserved quantities. We acknowledge the uncertainties in these observed relations, in particular, due to the limits of spherically symmetric simulations. It would be interesting to see if one can understand the condensation of the halo-soliton system in the spirit of \cite{Levkov_2018, dmitriev2024selfsimilar}, however many of our conclusions are independent of the exact form of the halo-soliton equilibrium as long as one believes there exists such a thing. If the relative abundance of the two species is the same for all galaxies, then the theoretical prediction of $\rho_c\propto 1/R_c^4$ remains. In this case, multi-field models do not provide an improved fit to the data. However, quite interestingly, if the relative abundances systematically change from lighter galaxies to heavier galaxies, then more general values of $\beta$ can be achieved. 

By parameterizing, the scaling of the relative fraction as $F\propto M^\alpha$, with $\alpha$ some power, we considered the more general prediction for $\beta$. There turns out to be significant dependence on the ratio of particle masses. An approximate formula for the value of $\beta$ as a function of $\alpha$ and the particle mass ratio $m_2/m_1$ is given in Eq.~(\ref{eq:beta}). For more discussion see the Appendix. This formula suggests that the observed value of $\beta=1.3$ is generally achieved if the value of $\alpha$ is negative. In other words, one consistently needs the more massive galaxies to be populated by a relatively larger abundance of the lighter species of particles. Some examples are given in Figure \ref{fig:correctscalings}. The ratio of particle masses does need to be quite significant; at least a ratio of 1000 or more, to cover the observed spread of galaxies (this point was nicely made in Ref.~\cite{Guo_2021}). For this rather large hierarchy of particle masses, our numerics currently do not have enough precision to determine $\alpha$ with high confidence. However, pinning down $\alpha$ for such a ratio of masses is a natural step for future work.

So for multi-field ULDM to have some improved fit to the data, one needs a non-trivial scaling of relative abundances among galaxies. Curiously, there is some evidence in the literature that during the formation process of a halo, the lighter species can cause tidal stripping of the heavier species \cite{Huang_2023,luu2023nested}. However, how this affects the final scaling relation of relative abundances deserves investigation. Even if the effect is small, we expect that this could already be sufficient for the most interesting mass ratios of $m_2/m_1 < 10^{-3}$, as Eq.~\eqref{eq:beta} seems to suggest that for smaller mass ratios the necessary $\alpha$ approaches closer to $0$. Exploring whether a small negative value of $\alpha$ can emerge dynamically is an important avenue for future research.

If indeed multi-field wave Dark Matter is the way to solve the scaling issue of Fig.~\ref{fig:obspcvsRc}, we can derive two other predictions that serve as consistency checks. In particular, we can get an estimate of the ratio of masses of the most massive and lightest galaxies in our dataset, which for $m_2/m_1 < 10^{-3}$ seems consistent with the dataset under consideration. Furthermore, we predict that the total mass of galaxies is positively correlated to the dilutedness of the core. In particular, the more massive the galaxy, the larger the core is in spatial extent, and the smaller its central density. Interestingly, this is different from the prediction of single-field ULDM which through relation Eq.~\eqref{eq:corehalomassrel} predicts the opposite and these correlations can thus be used to distinguish the two. These aspects of the model are highlighted in appendices~\ref{sec:appendixB} and \ref{sec:appendixC}.

Future work can involve 3-dimensional simulations, rather than the spherically symmetric simulations performed here. 
%We have explored a range of random initial conditions, but a more general and systematic exploration would be useful to confirm these results.
Furthermore, more refined simulations would be useful to definitively establish the core-halo relations that we have seen here. Another possible option is to go beyond two-fields to $N$-fields. While we think the two-field case is indicative of the trend for larger $N$, relative to the single species case, it is worth exploring in more detail. In fact, large $N$ has a phenomenological motivation: ULDM single models lead to significant density fluctuations that can cause heating and redistribution of stars in a way that is incompatible with data \cite{dalal2022fuzzy}. However, for large $N$, these density fluctuations are reduced as $\sim1/\sqrt{N}$ due to a random walk; hence the heating effects should become small. Then the core-halo and core-radius relations become critical in evaluating the ULDM proposal, as we have explored here.

To conclude, extensions of ULDM with more light degrees of freedom have many interesting phenomenological features originating from the properties of their cores. There is reason to believe that these models can resolve an issue with ULDM highlighted by one of us in \cite{Deng_2018} if there are dynamical ways in which the relative abundance of the two species can universally depend on the total halo mass. We showed that if this is the case the model makes two other phenomenological predictions, in particular about the ratio of the most massive and lightest galaxies, and the correlation between the dilutedness of the core and the halo mass. These serve both as a crosscheck and an interesting avenue for future investigation. For masses $m_2/m_1 < 10^{-3}$ all of these crosschecks agree with experimental data, although with some healthy amount of uncertainty. It is therefore clear that these models of Multi-Field ULDM warrant further research to remove some of these uncertainties.

\section{Acknowledgements}
We thank Rodrigo Vicente and Demao Kong for various useful discussions. We are also thankful to Davi Rodrigues for providing the empirical data we present in Appendix \ref{sec:appendixC} and which is first presented in Refs.~\cite{Rodrigues_2014,Rodrigues_2017}. We aslo thank him for useful remarks on the draft of this paper. M.~P.~H. is supported in part by National Science Foundation grant PHY-2310572. F.~D. acknowledges the support from the Departament de Recerca i Universitats from Generalitat de Catalunya
to the Grup de Recerca 00649 (Codi: 2021 SGR 00649) and funding from the ESF under the program Ayudas predoctorales of the Ministerio de Ciencia e Innovación PRE2020-094420.

\appendix

%\section{The Non-Relativistic Limit of Scalar Fields}
%\label{sec:appendixA}

\section{Ground State Gravitational Solitons}
\label{sec:appendixA}
\begin{figure}[t!]
    \centering
    \includegraphics[width = \textwidth]{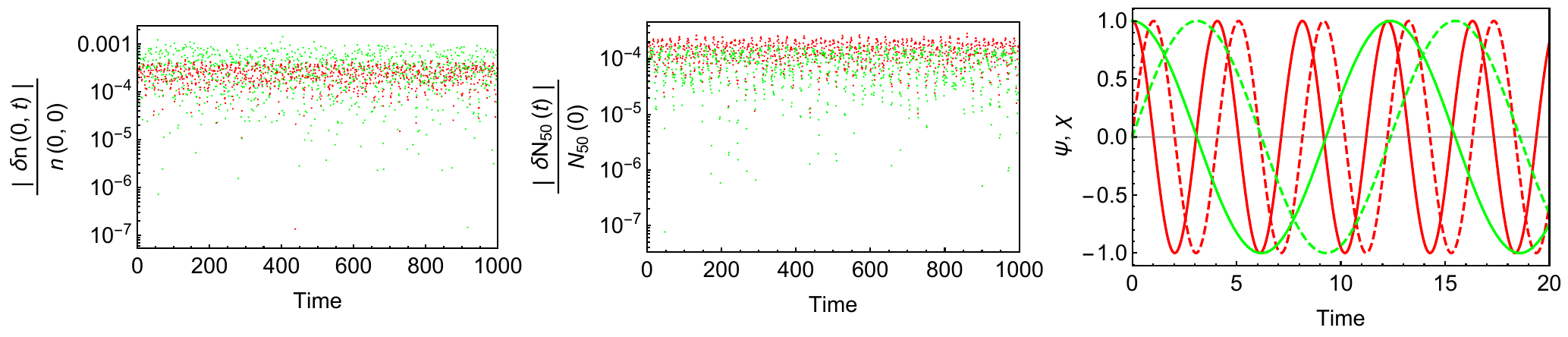}
    \caption{Some of the dynamical properties of the benchmark soliton for $m_2/m_1 = 1/2$ (the magenta profile in Fig.~\ref{fig:groundstatesolutionsm20d5}). \textit{Left:} The relative variation of the central density as defined in \eqref{eq:centraldensconserv} for the heavy field $\psi$ (red) and the light field $\chi$ (green). \textit{Middle:} The relative variation of the density profile as defined through \eqref{eq:numbconserv} for the heavy field $\psi$ (red) and the light field $\chi$ (green). \textit{Right:} The real (full) and imaginary (dashed) parts of the heavy field $\psi$ (red) and the light field $\chi$ (green) at the origin. As expected the real and imaginary parts oscillate out-of-phase with a phase shift of $\pi/2$.}
    \label{fig:staticsolm20d5}
\end{figure}

\begin{figure}[t!]
    \centering
    \includegraphics[width = \textwidth]{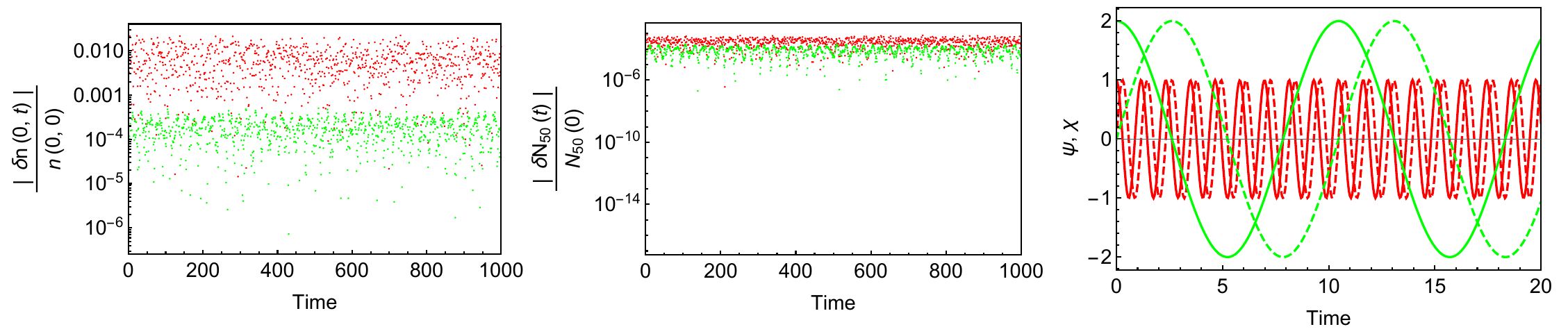}
    \caption{Some of the dynamical properties of the benchmark soliton for $m_2/m_1 = 1/5$ (the cyan profile in bottom panel of Fig.~\ref{fig:groundstatesolutionsm20d5}). \textit{Left:} The relative variation of the central density as defined in \eqref{eq:centraldensconserv} for the heavy field $\psi$ (red) and the light field $\chi$ (green). \textit{Middle:} The relative variation of the density profile as defined through \eqref{eq:numbconserv} for the heavy field $\psi$ (red) and the light field $\chi$ (green). \textit{Right:} The real (full) and imaginary (dashed) parts of the heavy field $\psi$ (red) and the light field $\chi$ (green) at the origin. As expected the real and imaginary parts oscillate out-of-phase with a phase shift of $\pi/2$.}
    \label{fig:staticsolm20d2}
\end{figure}

It is important to check whether the method highlighted in section~\ref{sec:staticsols} actually yields ground state solutions of the two-field SP system. In order to check this explicitly we ran simulations where the initial conditions were set to be exactly one of the zero-node solutions we found using the algorithm of Sec.~\ref{sec:staticsols}. A proper ground state of the system should have a constant density profile for many periods of oscillations in complex fieldspace. We found that this is the case for all our zero-node solutions. To check these properties it is useful to define the following quantities
\beq
\frac{|\delta n_\psi (0, t)|}{n_\psi (0, 0)} = \frac{|1 - |\psi (0, t)|^2|}{|\psi (0, t)|^2} \quad \textrm{and} \quad \frac{|\delta n_\chi (0, t)|}{n_\chi (0, 0)} = \frac{|1 - |\chi (0, t)|^2|}{|\chi (0, t)|^2}
\label{eq:centraldensconserv}
\eeq

\beq
\frac{|\delta N_{50, \psi} (t)|}{N_{50, \psi} (0)} = \frac{|1 -  N_{50, \psi} (t)|}{N_{50, \psi} (0)} \quad \textrm{and} \quad \frac{|\delta N_{50, \chi} (t)|}{N_{50, \chi} (0)} =  \frac{|1 -  N_{50, \chi} (t)|}{N_{50, \chi} (0)}
\label{eq:numbconserv}
\eeq
Where we define $N_{50}$ as the total amount of particles in each species, within the radius that initially contains $50\%$ of the total amount of particles of that species.
In Figs.~\ref{fig:staticsolm20d5} and ~\ref{fig:staticsolm20d2} we show this for two benchmark cases in which the two fields have comparable mass content and central density (in particular, the magenta ($m_2/m_1 = 1/2$) and cyan ($m_2/m_1 = 1/5$) profiles of Fig.~\ref{fig:groundstatesolutionsm20d5}). 

As is expected of ground state solitons the quantities defined in Eqs.~\eqref{eq:centraldensconserv} and \eqref{eq:numbconserv} are conserved for these solutions up to at most one part in one hundred (which is also somewhat dependent on the resolution of the simulation). This indicates that both the central densities and overall density profile of the two species stay constant. Finally, we observe that the real and imaginary parts of the two fields oscillate out of phase (with a phase-shift of $\pi/2$) as is to be expected of an eigenstate of the system. Interestingly the periods of oscillation need not be the same and can differ widely. These results for the benchmark cases, where both species have comparable total mass give us confidence in concluding that our algorithm is sufficient to find the ground state solitons of the two-field SP system.

\section{Understanding the scaling parameter $\beta$}
\label{sec:appendixB}
\begin{figure}
    \centering
    \includegraphics[width = 0.8\textwidth]{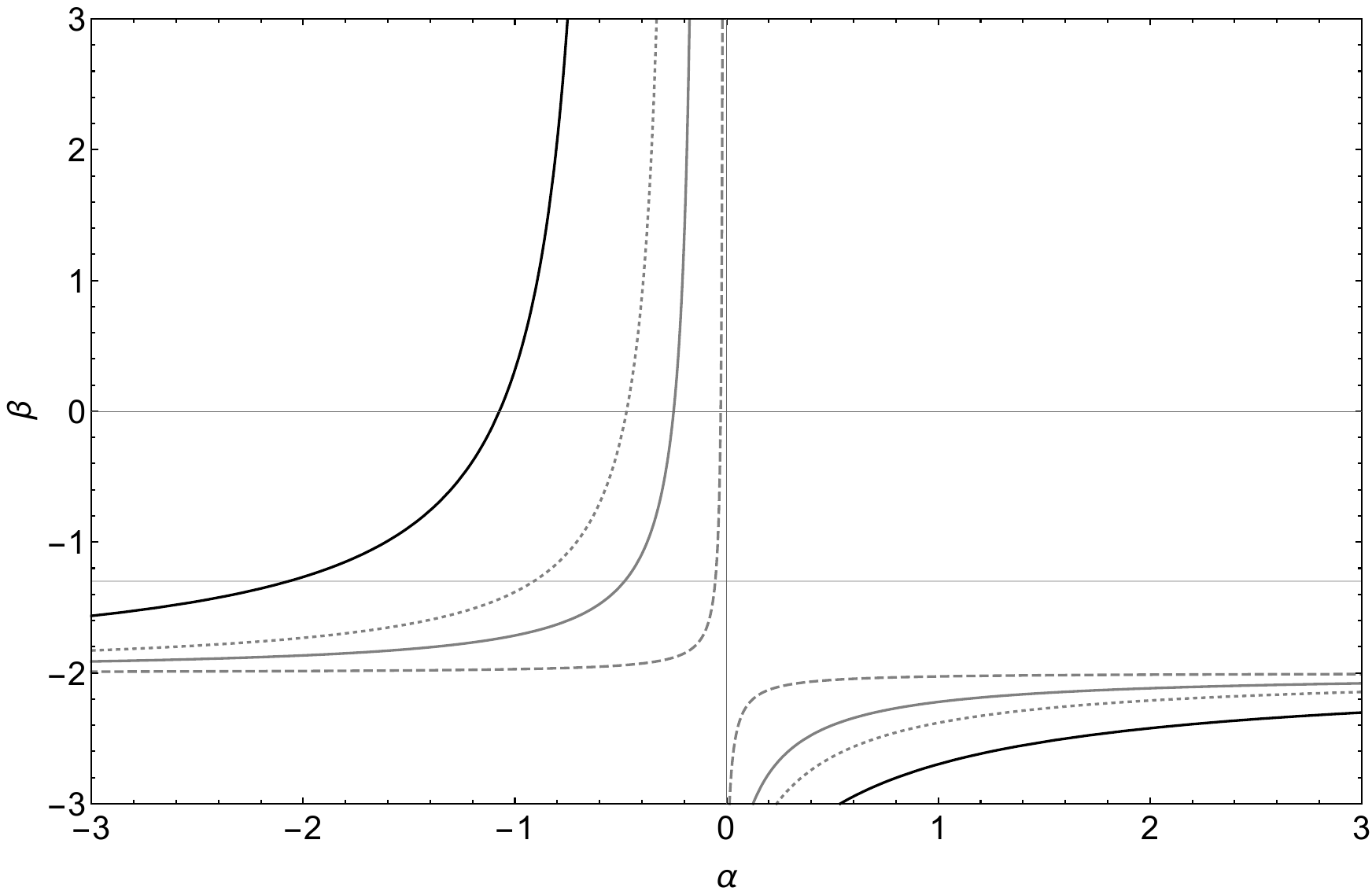}
    \caption{The expected value of $\beta$ depending on $\alpha$ for two different mass ratios $m_2/m_1 = 1/5$ (black) and $m_2/m_1 = 1/1000$ (grey). The two branches of the solution approach $\beta = -2$ from different directions and intersect the correct value of $\beta = -1.3$ when $\alpha < 0$. Due to the uncertainty in the exact equilibrium state of the halo-soliton system for the small mass ratio $m_2/m_1 = 1/1000$, we also plot the curves for two other values of $F_\chi/F_\psi$, increasing (dashed) and decreasing (dotted) it by a factor of $10$. We see that the necessary value of $\alpha$ does not depend too strongly on the exact equilibrium state of the halo-soliton system.}
    \label{fig:betavsalpha}
\end{figure}

It is possible to get a good understanding of what scalings typically emerge, given the dependence of $F\propto M^\alpha$ and the ratio of fundamental masses $m_2/m_1$. As a starting point, one can approximate the red dotted lines of Fig.~\ref{fig:scalingscosmo} as straight lines interpolating between the two extremal single-field scalings. In the single-field limit, we can immediately find the exact core solution by solving for the effective $\lambda$ in Eqs.~\eqref{eq:psitrans} and \eqref{eq:phitrans}, which through Eqs.~\eqref{eq:thermorel} must have $\lambda^2 \propto |E_h|/M_h$. If the halo collapsed from a homogeneous over-density, $E_h \propto M_h^{5/3}$. Thus, finally we conclude, for $F \propto M^\alpha$, that $\lambda \propto F^{1/3\alpha}$. The scaling $\rho_c \propto R_c^\beta$ that emerges is then
\beq
\beta \approx \frac{\log(\rho_{c, \chi}) - \log(\rho_{c, \psi})}{\log(R_{c, \chi}) - \log(R_{c, \psi})} = \frac{\log(\left(\frac{m_2}{m_1}\right)^2 \left(\frac{\lambda_\chi}{\lambda_\psi}\right)^4)}{\log(\left(\frac{m_2}{m_1}\right)^{-1} \left(\frac{\lambda_\chi}{\lambda_\psi}\right)^{-1})}= \frac{\log(\left(\frac{m_2}{m_1}\right)^2 \left(\frac{F_\chi}{F_\psi}\right)^{4/3\alpha})}{\log(\left(\frac{m_2}{m_1}\right)^{-1} \left(\frac{F_\chi}{F_\psi}\right)^{-1/3\alpha})}
\label{eq:scalingderivation}
\eeq
Where the subscript $\psi$ and $\chi$ should be understood as the values of $F$ and $\lambda$ for which the respective field completely dominates the core soliton and thus its core density and radius. More intuitively one can understand the values of $F_\chi$ and $F_\psi$ as follows: if $F > F_\psi$ we assume the stable equilibrium state is such, that the core (at least to the level that is observationally distinguishable) is completely dominated by the heavier $\psi$ field. The opposite is true whenever $F < F_\chi$, when the lighter $\chi$ field determines the properties of the core. Interesting scalings thus emerge whenever $F_\chi < F < F_\psi$ as can be seen in Fig.~\ref{fig:scalingscosmo} as the interpolating region in between the two asymptotes. Note that this line of reasoning holds regardless of what the actual equilibrium state of any given halo is, as long as one assumes that an equilibrium state exists and that one should have $F_\psi > F_\chi$. Although we are determining $F_\psi$ and $F_\chi$ using the algorithm of Sec.~\ref{sec:halosoliton}, our results should be quite general, regardless of the validity of Eqs.~\eqref{eq:thermorel}. Inspecting Eq.~\eqref{eq:scalingderivation} we see that the exact scaling that emerges depends on $m_2/m_1$, $\alpha$, and finally $F_\chi/F_\psi$. Interestingly this reasoning allows us to make another prediction for the model, namely, the ratio of the most massive and lightest galaxies in the ``scaling'' regime is given by
\beq
\frac{M_{heavy}}{M_{light}} = \left(\frac{F_\chi}{F_\psi}\right)^{\frac{1}{\alpha}}
\label{eq:MassRatGal}
\eeq
We can then solve for $\frac{F_\chi}{F_\psi}$ in Eq.~\eqref{eq:scalingderivation} in terms of $\beta$ and $\alpha$ en plug this back into Eq.~\eqref{eq:MassRatGal}. Interestingly $\alpha$ drops out to obtain
\beq
\frac{M_{heavy}}{M_{light}} = \left(\frac{m_2}{m_1}\right)^{\frac{-3(\beta + 2)}{4 + \beta}}
\label{eq:MassRatGalfinal}
\eeq
This is a crosscheck, independent of the specific halo-soliton equilibrium state, that we will come back to later since any Dark Matter model can not have this ratio be either too small or too large.
Eq.~\ref{eq:scalingderivation} has two branches which asymptote to $\beta = -2$ in the limit $|\alpha| \rightarrow \infty$, independent of $m_2/m_1$. The two branches approach the asymptote from different directions, and it is thus possible to obtain any value of the scaling parameter $\beta$, given the right $\alpha$. Although the solution for $\alpha = 0$ is not well defined, the limit of $\alpha \rightarrow 0$ gives $\beta = -4$ as expected. To solve the tension in Fig.~\ref{fig:obspcvsRc} in any meaningful way it is thus necessary that the ratio of abundances of Dark Matter species scales with the total mass of the galaxy in a universal way. However, to get a general estimate for $\beta$, we need an approximation for $F_\chi/F_\psi$, which could in general depend on $m_2/m_1$. Although it is very hard to predict a priori what the ratio of these quantities should be for any given $m_2/m_1$, we can get an idea for the trend of $F_\chi/F_\psi$ by investigating the results in Fig.~\ref{fig:scalingscosmo} and analogous results for different mass ratios. In particular, we find the scaling parameter $\beta$ explicitly using the method highlighted in Sec.~\ref{sec:cosm} at fixed $\alpha = 1$, for four different mass ratios $m_2/m_1 = 1/5, 2/5, 1/2, 4/5$. We then solve for $F_\chi/F_\psi$ in Eq.~\eqref{eq:scalingderivation}. We find that the value of $F_\chi/F_\psi$ is slightly increasing as we lower the ratio of $m_2/m_1$. Admittedly there is quite some uncertainty in these conclusions and our results are also consistent with $F_\chi/F_\psi \approx 0.075$ (constant) for smaller values of $m_2/m_1$, which is what one could expect looking at the dependency of the ratio of total mass in the solitons themselves on the central field values in Fig.~\ref{fig:Fvsf}, which seem to follow simple power laws. Also, $F_\chi/F_\psi < 1$ is required by consistency ($F_\chi > F_\psi$ would indicate that the heavier species dominates the soliton only when there is a smaller abundance of the heavier particle present, which, although intriguing, we exclude here). We thus assume $F_\chi/F_\psi = 0.075$ to obtain
\beq
\beta \approx \frac{\log(0.075^{4/3\alpha}) + \log(\left(\frac{m_2}{m_1}\right)^{2})}{\log(0.075^{-1/3\alpha}) + \log(\left(\frac{m_2}{m_1}\right)^{-1})}
\label{eq:empericrelationbeta}
\eeq
In Fig.~\ref{fig:betavsalpha} we plot the dependence of $\beta$ on $\alpha$ for two different ratios of $m_2/m_1$. The two branches of the solution are visible, intersecting the correct value of $\beta = 1.3$ for a single value of $\alpha$. This value is represented as the gridline in Fig.~\ref{fig:betavsalpha}.
Using Eq.~\eqref{eq:empericrelationbeta} we can estimate the necessary $\alpha$ to reproduce $\beta = -1.3$ for any $m_2/m_1$. A particularly interesting case has (as first pointed out in \cite{Guo_2021}) $m_2/m_1 = 10^{-3}$ since for $m_1 \sim \SI{e-21}{\eV}$ and $m_2 \sim \SI{e-24}{\eV}$ all the datapoints in Fig.~\ref{fig:obspcvsRc} can be fitted by a groundstate soliton. Essentially, all the data falls in between the two extremal single-field lines. With our assumption that $F_\chi/F_\psi = 0.075=$ Constant, independent of the exact value of $m_2/m_1$, we find that $\alpha \approx -0.5$ to reproduce the correct scaling. However, as there is quite some uncertainty in the exact value of $F_\chi/F_\psi$, we also investigated what would be needed if we increase and decrease the assumed value of $F_\psi/F_\chi$ by a factor of $10$. It turns out that the dependence of this exact value is not too strong and that $\alpha$ fluctuates between $-1$ and $-0.1$. The range is shown with the three gray lines in Fig.~\ref{fig:betavsalpha}. Interestingly, for this ratio of $m_2/m_1$, the ratio of the heaviest and lightest galaxies, for $\beta = -1.3$, through Eq.~\eqref{eq:MassRatGalfinal} is $M_{heavy}/M_{light} = 215 \sim O(10^2 - 10^3)$, conforming to the dataset under consideration (see also Appendix~\ref{sec:appendixC}). This result is independent of $F_\chi/F_\psi$ and serves as a good sanity check for the validity of the two-field model as the solution of the scaling problem. While the exact value of $\alpha$ needed is somewhat approximate due to the uncertainty in $F_\chi/F_\psi$ and the exact equilibrium the soliton-halo system reaches in this two-field model we obtain a clear trend of what would be needed in multi-component ULDM models to solve the tension highlighted in Fig.~\ref{fig:obspcvsRc}. In particular a small negative value of $\alpha$.

\section{Relation between Soliton Properties and Halo Mass}
\label{sec:appendixC}
\begin{figure}[h]
    \centering
    \includegraphics[width = \textwidth]{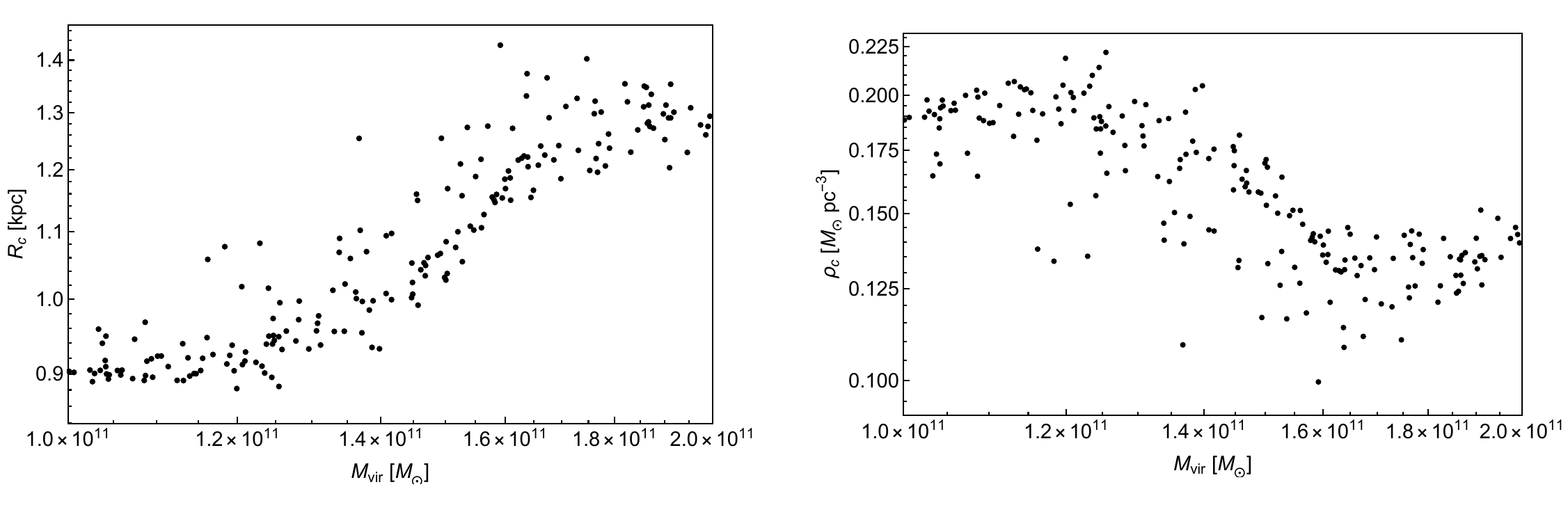}
    \caption{The core density and radius versus the total virial mass for the galaxies of in the top right corner of Fig.~\ref{fig:correctscalings}. Since here we have $\alpha < 0$ the correlation between galaxy mass and core central density is negative, while the opposite is true for the core radius and galaxy mass. This agrees with the weak correlations of Fig.~\ref{fig:corevsmvir} and we want to stress that we expect this to hold in general for two-field ULDM models if $\beta = -1.3$, since this is only possible on the $\alpha < 0$ branch of Eq.~\eqref{eq:beta}.}
    \label{fig:corevsmvirmock}
\end{figure}

\begin{figure}[h!]
    \centering
    \includegraphics[width = \textwidth]{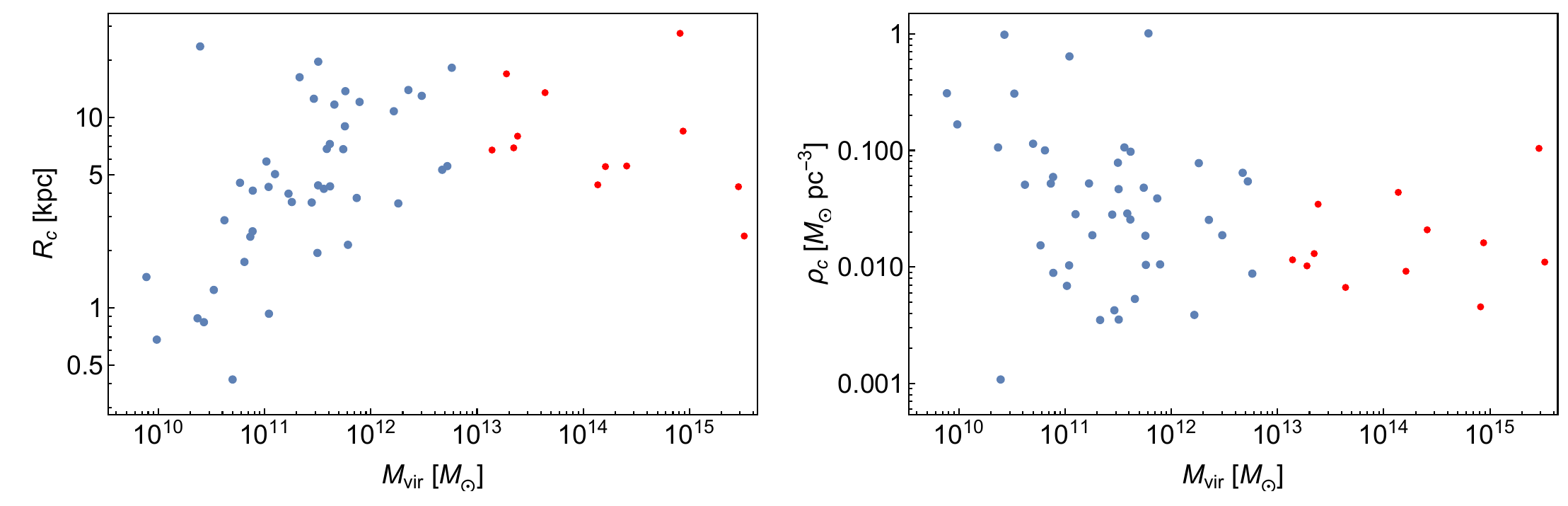}
    \caption{The core density and radius versus the total virial mass for the galaxies of Fig.~\ref{fig:obspcvsRc}. The data shows a hint of some possible correlation, especially if we exclude galaxies that we consider to have unphysical masses 
    of $M_{vir} > 10^{13} M_\odot$ (which are plotted in red here). Even with this exclusion, however, the correlation is not very strong. Nonetheless, a positive correlation between the core radius and virial mass, and a negative correlation between the core density and virial mass can be seen. This contradicts the scaling of Eq.~\eqref{eq:corehalomassrel}, but interestingly agrees with the mock galaxies of Fig.~\ref{fig:correctscalings}.}
    \label{fig:corevsmvir}
\end{figure}

There is another hidden prediction in the generated datasets of Sec.~\ref{sec:cosm}. In particular, it relates the total mass of each galaxy to the central core properties. To zeroth-order it is simple to see how these quantities are correlated in the mock galaxy sets we generated. In particular, with the assumptions taken, the energy per unit mass of the halo is positively correlated to the halo mass: $E_h/M_h \propto M_h^{2/3}$. If we were dealing with a single-field system and the energy per unit mass in the halo were the same as in the soliton (as we're assuming here), we would immediately conclude that the mass of the soliton scales as $M_s \propto \lambda M_0 = M_h^{1/3} M_0$, as suggested in Eq.~\eqref{eq:corehalomassrel}. At zeroth-order there then should be a positive correlation between the soliton mass and the halo mass. In the single-field limit, this translates into a positive correlation between core density and halo mass, and a negative correlation between core radius and halo mass. In the two field case these correlations (in the interpolating region in which both species are important in determining the properties of the core) depend entirely on which branch of Eq.~\eqref{eq:scalingderivation} the solution for $\beta$ is found. On the $\alpha \rightarrow \infty$ branch the correlations are the same as in the single-field case. However, on the opposite branch of $\alpha \rightarrow -\infty$ the correlations are reversed. We show in Fig. \ref{fig:corevsmvirmock} what the behavior is for the galaxies generated in Sec.~\ref{sec:cosm}, with $m_2/m_1 = 1/2$ and the correct scaling of $\beta = -1.3$ (in particular the top right graph of Fig.~\ref{fig:correctscalings}). Since here we have $\alpha < 0$ the correlation between galaxy mass and core central density is negative, while the opposite is true for the core radius and galaxy mass. It is important to note that the opposite would indeed be the case if the solution was found for $\alpha > 0$, agreeing with the single field limit.
We can compare this prediction with empirical data from \cite{Rodrigues_2014} and \cite{Rodrigues_2017}. Combining the two sources we obtain estimates of core properties and halo masses for 56 galaxies. The estimates are obtained by fitting rotation curves with different DM density profiles. In particular, the virial mass plotted here $M_{vir}$ corresponds to the mass of the best-fit NFW profile integrated up to $r_{200}$, defined through
\begin{equation}
    M_{NFW} (r_{200}) = 200 \frac{4 \pi}{3} r_{200}^3 \rho_{\mathrm{crit}}
\end{equation}
Where $\rho_{\mathrm{crit}}$ is the critical cosmological density. We want to note that the exact definition of the virial mass impacts the implied virial mass. In general, it is not easy to find a consistent definition of the mass of a galaxy, that is exempt from systematics. In particular, using different fits for the density profiles of galaxies will lead to a different implied virial mass. For example, in other works the Burkert profile was used to fit rotation curves and derive the mass of each galaxy \cite{Li_2020, Rodrigues_2023}, changing the results somewhat. The results are plotted in Figs.~\ref{fig:corevsmvir}.
Although the correlations are weak, the data seems to be in tension with single component ULDM models. One can alleviate this tension for $\alpha < 0$ (we defined $\alpha$ through $F = M_\psi/M_\chi \propto M^\alpha$, the dependence of the relative abundance of species on the total galaxy mass). In fact, one can reconcile both the tension highlighted here and that of Fig.~\ref{fig:obspcvsRc}, since for $\beta = -1.3$, $\alpha < 0$ necessarily. The data puts strain on the single-field models suggesting the scaling of Eq.~\eqref{eq:corehalomassrel}, which are not able to produce these correlations (if Eq.~\eqref{eq:corehalomassrel} holds). There is however large uncertainty in these conclusions, as the correlation in the experimental data is weak, and the definition of the virial mass is not exempt from systematics. A more comprehensive and complete analysis of the relation between the core properties and halo mass of various galaxies is an interesting avenue for future research.

\bibliography{multiuldm}

\end{document}